\documentclass[useAMS,usenatbib]{mn2e}

\usepackage{graphicx}
\usepackage{subfig}

\title[Evolution of disc-planet systems with an inclined binary companion]{Evolution of  a disc-planet system  with a  binary companion on an inclined orbit }
\author[M. Xiang-Gruess and J.C.B. Papaloizou]{M. Xiang-Gruess$^{1}$ \thanks{E-mail:
mx216@cam.ac.uk } and J.C.B. Papaloizou$^{1}$ \\
$^{1}$ DAMTP, University of Cambridge, Wilberforce Road, Cambridge CB3 0WA, UK\\
}
\begin{document}

\date{Accepted . Received ; }

\pagerange{\pageref{firstpage}--\pageref{lastpage}} \pubyear{2002}

\maketitle

\label{firstpage}

\begin{abstract}
We  study  orbital inclination  changes  associated with the precession  of 
a disc-planet system that occurs  through  gravitational interaction with
a  binary companion  on an inclined orbit. We  investigate  whether
  this  scenario  can account for   giant planets on close  orbits 
  highly inclined  to the stellar equatorial plane.  We   obtain  conditions 
 for  maintaining  approximate coplanarity and test them  with   SPH-simulations.   For  parameters of 
interest, the  system undergoes  approximate rigid body precession  with    modest 
 warping while the  planets migrate inwards.  Because of  pressure forces,  
disc self-gravity is not needed   to maintain the configuration.  We 
 consider a disc and single  planet for different  initial  inclinations
 of the binary orbit to the midplane of the   combined system and  a system of three planets  
 for which  migration leads to  dynamical instability  that reorders the
 planets. As the interaction  is dominated by  the time averaged
 quadrupole  component of the binary's  perturbing potential, results for  a
 circular  orbit can be scaled to apply to  eccentric orbits. 
The  system  responded adiabatically  when  changes to binary  orbital parameters 
 occurred  on  time scales exceeding the  orbital period.
Accordingly  inclination changes are  maintained under its slow removal. Thus
 the  scenario for generating high inclination  planetary orbits   studied here, is promising. 
\end{abstract}


\section{Introduction}

In the past few decades, the number of detected extrasolar planets around other main sequence stars has increased dramatically, so that  at the time of writing,  more than 1000 planets have been discovered.
Well studied  planet formation  scenarios  such as  core accretion  \citep{Miz1980, Pol1996} or
disc fragmentation \citep{May2002} involve the planet forming in a disc with the natural expectation
that the orbit should be  coplanar. However, Rossiter McLaughlin measurements  of close orbiting hot Jupiters
indicate that around $40\%$ have angular momentum vectors  significantly misaligned with  the
angular velocity vector of the central star \citep[e.g.][]{Tri2010, Alb2012}. As the stellar and disc angular velocities
are naturally expected to be aligned, this would imply an inclination of the planet orbit relative to the nascent protoplanetary disc.

This has been underlined by 
several recent studies of the interactions of planets with a range of initial orbital inclinations with respect to a
 disc \citep[e.g.][]{Cre2007, Mar2009, Bit2011}.
In these works the evolution of planets with masses  up to a maximum of  $1\rmn{M_J}$ with different initial eccentricities and relatively small
 initial inclinations up to a maximum of $15^\circ$ interacting with three-dimensional isothermal and radiative discs are considered.
 They showed damping for both eccentricity and inclination with the planets circularising  in the disc after a few hundred orbits.
Planets with higher inclinations have also been studied by e.g. \cite{ Rei2012}  and  \cite{Tey2013}.
A detailed survey of a range of planet masses and the full range of orbital inclinations interacting with a disc  has been presented by \cite{Xia2013} (hereafter Paper 1).
The timescale of realignment with the gas disc was found to be  comparable with the disc  lifetime for very high inclinations $ >60^{\circ}$ and planet masses of  one Jupiter mass.
 For smaller initial relative inclinations  and/or  larger planet masses,  the timescale for  realignment is shorter than the disc lifetime.
These results taken together imply that there is  a need for an explanation for the origin  of  misaligned giant planets,
if it is assumed that the discs angular momentum vector is always  aligned with the spin axis of the central star.

Several  scenarios have been proposed to explain the origin of 
misalignments between the planetary orbital angular  momentum vector  and the stellar rotation  axis  as well as produce close in giant planets.
The first involves excitation  of very high eccentricities, either through  the Lidov-Kozai
effect induced by the  interaction with a distant companion \citep[e.g.][]{Fab2007, Wu2007},
or through planet-planet scattering or chaotic interactions \citep[e.g.][]{Weid1996,Rasio1996,Pap2001,Nag2008}.
When the planet attains a small enough pericentre distance, 
this is then followed by orbital circularization due to tidal interaction with the central star, leaving the planet on a close inclined  circular orbit.

A method for producing high disc inclinations with respect to the stellar equator
has also been indicated by \cite{Tho2003}, who  have studied the evolution of two giant planets in resonance,
adopting an approximate analytical expression for the influence of a gas disc in producing  orbital migration.
Their calculations suggest that resonant inclination excitation can occur when  the eccentricity of the inner planet reaches a threshold value  $\sim  0.6$.
 Inclinations gained by the resonant pair of planets can reach values up to $\sim 60^\circ$. 

However in  a recent study,   \citet{Daw2013} find that large eccentricities,  that are expected to  be associated 
with such interactions for  orbits too wide to be affected by stellar tides,   are seen  mostly
in metal rich systems.
Accordingly they conclude  that  gentle  disc migration and planet-planet scattering must  both operate during the early evolution of giant planets.
Only systems packed with giant planets, which most easily form around metal rich stars, can produce large eccentricities through scattering.
They also note a lack of a correlation between spin-orbit misalignment and metallicity that  may indicate the operation of a mechanism 
that can cause misalignment of the disc while simultaneously  allowing disc migration of giant planets into close orbits.


An explanation for the origin of  misaligned planets, that is consistent  with formation in  and inward migration driven by a disc may be
  connected with the possibility that the orientation of the disc changes,  possibly due  either to  a lack of orbital  alignment of the source material or to gravitational encounters
with passing stars. This may occur   either before or after  planet formation  \citep[e.g.][]{Bat2010,Thi2011}.

\cite{Thi2011}  showed that the capture of gas from accretion envelopes by  a protoplanetary disc could cause it to become significantly misaligned with respect to its original plane.
 Although no planets are involved in their simulations, it is of   interest to remark that  misaligned gas discs can easily be produced
 when  interactions  with the environment out of which they form are considered.
 In this regard we note that  protostars are not generally isolated but are  bound in stellar clusters where gravitational interactions  are common.

Recently, the system Kepler-56 has been studied by \cite{Hub2013} who used asteroseismological methods to determine the obliquity in that system. Kepler-56 has  two known transiting planets. \cite{Hub2013} found that the planets have angular momentum vectors  strongly inclined  with  respect to the central stellar spin axis.
Taking into account the large observed obliquities,  dynamical simulations with the two known planets  failed to produce a stable non coplanar  planetary system.  
The authors then studied an alternative scenario involving a distant companion and were able to show in dynamical simulations that they succeeded in producing the architecture of the Kepler-56 system including  coplanarity of the planetary orbits  and the high obliquity. Further observations are required to specify the mass of the companion which is crucial to determine  the origin of the misalignment in Kepler-56. In the case of a planetary companion, the misalignment is inferred to occur after the planets have  formed while in the case
 of a stellar companion, the misalignment is inferred to  occur while  the host gaseous disc is present.

 In this  context,  \cite{Bat2012} suggested that a binary companion  could be responsible for  producing  misalignments between stellar spin axes and planetary orbital planes
as a result of making  the protoplanetary disc precess.
The process envisaged the formation and disruption of wide binaries  in birth clusters. 
that dissolve on timescales up to a few tens of millions of years. 
The excitation of significant misalignment should  then occur within the same time scale.

We remark that the interaction between a binary companion and a disc could cause the disc to become warped. 
Notably, \cite{Ter2013} studied the effect of a warped disc on the orbit of a Jupiter mass  planet.
For a sufficiently warped disc, it was shown that  a  high orbital inclination  for the  planet  might be excited through direct gravitational interaction.

In view of the above discussion, it is important to investigate the response of a system composed of both a protoplanetary disc
and  one or more planets that can induce a gap and migrate within it,  under the gravitational perturbation of a binary companion in a misaligned orbit.
If  misalignment  with the stellar equatorial plane is induced during the late stages,  planets may reside in a cavity in the inner regions of the disc,
 relatively little further accretion of disc material onto the central star may take place optimising the final misalignment,  thus
making this an interesting case to study.

In this paper we consider the response of the  disc-planet system as a whole obtaining  conditions,  based on estimates of precessional torques,  for it to maintain approximate
coplanarity and test these against the results of numerical simulations.  From these  we are able to show that for appropriate parameters of intererest,
the whole system precesses approximately as a rigid body with the disc exhibiting only a modest amount of warping.
This does not   necessitate a self-gravitating disc as assumed by \citet{Bat2012} because the disc can communicate  more effectively through the action of pressure
 forces
\citep[eg.][]{Lar1996}  than through density waves as long as the Toomre stability parameter exceeds
unity,  a situation  expected  for the most part during the later stages of protoplanetary discs   \citep[see][for a detailed discussion]{Papl99}.
Even if the outer parts of the disc are maintained close to marginal stability, effects due to self-gravity and pressure are  expected to be comparable for global warps.
Accordingly we make the simplification of neglecting the disc self-gravity which results in a very significant reduction in computational resource requirements.
 We also test how the  response adjusts to changes  in the orbital parameters of the binary finding that
this is essentially adiabatic even when the changes occur on a timescale comparable to the binary orbital period.

It has been shown that Smoothed particle hydrodynamics (SPH) simulations are capable   of being applied to the problem of planet-disc interaction
\citep[e.g.][]{Sch2004,Val2006} though in comparison to grid-based simulation methods, they are generally more diffusive
and incur greater computational expense.
But in contrast to grid-based methods, as they adopt a Lagrangian approach,
SPH simulations can be readily used to simulate a gas disc  with a free boundary   that undergoes a large amount of movement  in three dimensions
with the disc having the freedom to change its shape at will. As it is suitable for a system precessing
with high mutual inclinations, we adopt this approach  here.

In Section \ref{sec:sim_details}, we   describe our simulation technique,    
giving details  of the modified locally isothermal equation of state we adopted in Section \ref{sec:eos}
and details  of the smoothing length and artificial viscosity prescriptions in Section \ref{sec:artv}.
In Section \ref{sec:IC}, we describe the general setup for  the disc,  the  planets, the central star and the binary companion.
In Section \ref{sec:theory}, we give a brief theoretical overview of the  global interaction between the planet-disc system and a   binary star
with orbital plane  misaligned with the  disc midplane.    
Estimates of the precessional torques that can
act  between different components in the system are given   in   appendix \ref{ap:precession}.
If the putative precession frequencies they could produce, significantly exceed the putative precession frequency of  the entire system
calculated assuming that it rotates as a rigid body, these torques can be utilised to maintain alignment.  
Precessional  Torques acting between a disc and a planet in a gap/inner cavity as well as between two planets without  a disc are considered.

We go on to present  the results of our simulations   in  Section \ref{sec:1pl}.
Simulations with a disc together with a single migrating giant planet and a binary companion
in an inclined orbit  are discussed in 
Section \ref{sec:1p2}. These  systems are found to maintain approximate alignment 
consistently with  estimated magnitudes of internal precessional  torques.
The adiabatic response to changes in the binary orbit is discussed in
Section \ref{Adi}.
A simulation with a disc, misaligned binary orbit and a three planet system is described in 
Section \ref{sec:3pl}. The disc-planet system  is seen to maintain approximate coplanarity, 
as a result of disc-planet torques, as  in the single planet case,
even though there is a phase of dynamical instability.
Finally, we sumarize and discuss our results in Section \ref{sec:concl}.

\section{Simulation details} \label{sec:sim_details}

We  have performed simulations using a modified version of the publicly available code GADGET-2  \citet{Spr2005}. 
GADGET-2 is a hybrid N-body/SPH code capable of modelling both
fluid and distinct  fixed or  orbiting  massive bodies. In our case the central star, of mass $M_*,$  is fixed while the planets, of masses $M_{p,i}, i = 1 , 2 ..$
 and the binary companion, of mass $M_B$ orbit as distinct massive bodies.
We adopt spherical polar coordinates $(r, \theta,\phi)$ with origin at the centre of mass of the central star.  The associated Cartesian coordinates $(x,y,z)$
are such that the $(x,y)$ plane coincides with the initial midplane of the disc.

\noindent  The gaseous disc is represented by SPH particles. An  important issue  for N-body/SPH simulations is the choice of the gravitational softening lengths.
In our simulations, the massive bodies (i.e. central star, companion and the planets) are unsoftened when interacting with each other,
allowing them to undergo realistic close encounters.

The total unsoftened gravitational potential $\Psi$ at a position $\bf{r}$ is given by \citep[see also][]{Pap1995, Lar1996}.
\begin{eqnarray}
 \Psi(\bf{r})&=& -\frac{\rmn{G M_*} }{|\bf{r}|} - \frac{\rmn{G M_B} }{|\bf{r}-\bf{D}|} + \frac{\rmn{G M_B} \bf{r}\cdot \bf{D} }{|\bf{D}|^3}\nonumber\\
&&- \sum_i \left(\frac{\rmn{G M_{p,i}} }{|{\bf{r}}-{\bf{r}}_{p,i}|}  - \frac{\rmn{G M_{p,i}}{ \bf{r}}\cdot {\bf{r}}_{p,i} }{|{\bf{r}}_{p,i}|^3}\right).
\label{Pot}\end{eqnarray}
Contributions from the central star, the binary companion of mass, $M_B,$
with  position  ${\bf{D}},$ together with
an arbitrary number of planets  with masses, $M_{p,i},$ and position
vectors ${\bf{r}}_{p,i},$  that are summed over,  are included.
Because  the origin of the coordinate system  moves with  the central star,
the  well known  indirect terms accounting for the acceleration of the coordinate system are present in equation (\ref{Pot}).

For the  computation  of the gravitational interaction between the
massive bodies  and the SPH particles,  the potential was softened following the   method of \citet{Spr2005}.
This was implemented with  fixed  softening lengths
$\varepsilon_*=\varepsilon_B=0.4\ \rmn{AU}$ and $\varepsilon_{p}=0.1\ \rmn{AU}$ for the central star, and the binary companion 
and  planets, respectively.

The disc acts  
gravitationally on the other bodies, but its self-gravity was neglected as this 
is not expected to play a significant role
for protoplanetary discs of the mass we consider.
For the computation of the gravitational force between  the disc and  the planets, we included  all SPH particles
in order to enable an accurate calculation when applying the tree algorithm.

The stars and the planets are allowed to  accrete gas particles that approach them very closely. We  followed  the procedure  of \cite{Bat1995} (see also Paper 1). 
This was applied such that for the stars and planets, the outer accretion radii were fixed during the simulation to be 
$R_{accr,*}=R_{accr,B}=7\times 10^{11}\ \rmn{cm}$, and $R_{accr, p}=7\times 10^{10}\ \rmn{cm}$, respectively.
The inner accretion radii were taken to be 0.5 of the outer accretion radii.
In practice, for the equations of state we used, the accretion of gas particles was found to 
play only a minor role in  
our simulations, producing negligible change to the masses of the planet and both stars.

\subsection{Equation of state} \label{sec:eos}

As described in Paper 1, we adopt  a locally isothermal equation of staete (EOS) incorporating  the modification of  \cite{Pep2008}. The sound speed $c_s$ is obtained from
\begin{eqnarray}
c_s =\frac{h_s r_s h_p r_p}{\left[ (h_sr_s)^n + (h_p r_p)^n \right]^{1/n}} \sqrt{\Omega_s^2 + \Omega_p^2}\ , \label{eq:Pep_mod}
\end{eqnarray}
with $n=3.5$. Here   $r_s=|\bf{r}|$ and $r_p=|{\bf{r}} -{\bf{r}}_{p,i}|$  are the distances to the central star and the nearest planet respectively.
 Here 
 $h_s=H/r_s$   is  the disc aspect ratio  with  $H$  being the circumstellar  disc scale height.   The angular velocities in the circumstellar and circumplanetary discs
are   $\Omega_s$ and $\Omega_p$ respectively.
We evaluate these assuming  Keplerian   circular orbits, thus  they are given by
\begin{equation}
 \Omega_s=\sqrt{\frac{\rmn{G} M_*}{r_s^3}}\hspace{2mm}    {\rm and} \hspace{2mm}  
 \Omega_p=\sqrt{\frac{\rmn{G} M_{p,i}}{r_p^3}}\hspace{2mm} {\rm respectively.}
\end{equation}

For  $r_s \leq 10\ \rmn{AU}$, we adopt  $h_s = 0.05.$
For simulations with a binary companion in an orbit inclined to the disc, it was found that  for high initial inclinations $ i_B \ge 60\,^{\circ},$
expansion of the outer regions of the disc could result in a small number of particles  being induced to interact strongly with
the companion. In order to avoid numerical problems arising from this, we adopted the following procedure.
For the outer parts with  $20\ \rmn{AU} > r_s >10\ \rmn{AU}$, we applied   a linear decrease of $h_s$ to $0.03.$ 
For $r\geq 20\ \rmn{AU}$, we adopted  a constant ratio $h_s = 0.03.$
We found that the application of  this procedure did not affect results for smaller values of $i_B < 60^{\circ}.$
Accordingly it was not adopted for the simulations of the multi-planet system  presented below.
Here we only study planet masses  $M_{p,i} \geq 1\ \rmn{M_J}$ for which the aspect ratio of the circumplanetary disc is taken to be  $h_p=0.6$ (see paper 1 for more discussion).

\subsection{Smoothing length and artificial viscosity} \label{sec:artv}

For our  SPH calculations, the smoothing length was adjusted so that the number of  nearest  neighbours  to any particle contained within a sphere of radius equal to the local smoothing length was  $40 \pm 5$. 
The pressure is given by  $p=\rho c_s^2$. 
Thus apart from  the vicinity of a planet, the temperature in the disc is $\propto r^{-1}.$ 
The artificial viscosity parameter $\alpha$ of GADGET-2 \citep[see equations (9) and (14) of][]{Spr2005} was taken to be $\alpha =0.5$. 

We remark that the artificial viscosity is modified by the application of a viscosity-limiter to reduce artificially induced  angular momentum transport in the presence of shear flows. This is especially important for the study of Keplerian discs.
Details are given in Paper 1. Also in Paper 1, we showed that runs with  $\alpha=0.5$  provided the best match to the analytic ring spreading solution with the \citet{Shak1973}
viscosity parameter  $\alpha_{SS}=0.02.$
SPH simulations accordingly model a viscous disc, with a viscosity that behaves like a conventional Navier Stokes viscosity in that context. However,
it is important to note the effective viscosity 
is likely to be flow dependent and so there   may not be a  particular form that applies to  general situations.

\section[]{Initial conditions} \label{sec:IC}

We study a system composed of a central star of one solar mass $\rmn{M_\odot}$, a binary companion of
mass  $M_B= 1\ \rmn{M_\odot}$, a gaseous disc and one or three planets initially embedded in the disc.
The disc is set up such that the angular momentum vector for all particles 
was in the same direction enabling a midplane for the disc to be defined.
 For simulations with a single planet, the planet mass was taken to be 
 $M_{p,1} =2\ \rmn{M_J}$ unless otherwise stated.
For the simulation with   three planets,  we chose planet masses in the range [0.4; 2 ] $\rmn{M_J}.$
The planets were initiated in circular  coplanar orbits  with zero inclination with  respect to the
 initial disc midplane in all cases.

The primary central star is  fixed at the origin of our coordinate system 
 while the binary star with separation $D$ orbits in a circle 
 about the primary central star with orbital angular velocity 
\begin{eqnarray}
 \omega_B= \sqrt{\frac{2 \rmn{G} \rmn{M}_\odot}{D^3}}\,
\end{eqnarray}
  with $D = |{\bf{D}}|.$
The gravitational effect of the disc and planets on the binary orbit is neglected.
In order to  avoid transients arising from an abrupt introduction of the binary star,
 we allow its mass  to increase linearly to 1 $\rmn{M_\odot}$ during the first 10 internal time units,
 with the internal time unit  being the orbital period at $a=5\ \rmn{AU}.$
Corresponding to this, the internal unit of length is taken to be  $5\ \rmn{AU}.$
These units of length and  time are  adopted  for all plots shown in this paper.

The particle distribution  was chosen to  model  a  disc with surface density profile  given by
\begin{eqnarray}
\Sigma=\Sigma_0 R^{-1/2}. \label{eq:sigma}
\end{eqnarray}
Here $\Sigma_0$ is a constant and $R$ is the radial coordinate of a point in the midplane.
This applied to
 the radial domain $[0, R_{out}-\delta]$, with  $R_{out}= 4a$ and $\delta =0.4 a$.
  A taper was applied  such that the surface density was set to decrease linearly to zero
  for $R$ in the  interval $[ R_{out}-\delta,  R_{out}+\delta].$
Disc  material thus  occupies the radial domain  $[0, R_{out}+\delta]$. 
 
\noindent The disc mass is given by
\begin{eqnarray}
M_D=2\pi \int_{R_{in}}^{R_{out}} \Sigma(r)r dr=\frac{4}{3} \pi \Sigma_0 R_{out}^{3/2}\ ,
\end{eqnarray}
which is used to determine $\Sigma_0.$
For the simulations presented here, we adopted $M_D = 10^{-2}\ \rmn{M}_{\odot}.$
The disc particles were set up in a state of pure Keplerian rotation according to
\begin{eqnarray}
 v_\varphi=\sqrt{r_s\frac{d\Phi_*}{dr_s}}\ ,
\end{eqnarray}
where $\Phi_*$ is the gravitational potential due to the central star.
In the innermost region around the central star, the disc properties (e.g. $\Omega_s$ and accordingly $c_s$) are modified by  the  softening
of the potential due to the central star.
For this reason simulations are terminated if migrating planets move inwards to the extent that the dynamics  starts to become  affected
by this region.


The  number of SPH particles   for most of  our simulations was taken to be  2$\times 10^5.$
This enabled a suite of simulations to be carried out for up to $600$ orbits in our dimensionless units.
This length of time was required  to  study the precession of the disc. 
Some of our simulations were also run with $10^5$ and $4\times 10^5$ particles in order to test the effect
of changing particle number. No significant changes occured. This siuation is the same as was indicated  for the
simulations presented in Paper 1.

For orbital separation of the binary star, we adopted  $D=100\ \rmn{AU}.$
This is distant enough to result in a regular almost rigid body precession of the disc, while  being close enough
to yield a short enough precession period that the disc precession could be studied.
We consider values of the inclination $i_B$ of the binary star with  respect to the $z=0$ plane in the reference frame of the host star
in the range $[0,\pi/2].$ In this context we remark that as the induced precession of the disc-planet system is consistent with
a secular interaction it should have little dependence on the  sense of rotation of the binary in its orbit, so that additional values of $i_B$ 
in the interval  $[\pi/2 , \pi]$  do not need to be considered.
We remark that  for $i_B$  approaching $\pi/2,$ the disc angular momentum vector may almost completely reverse relative to its initial direction
given a sufficiently long time.

\section{Influence of a  binary companion  on the  disc and planets} \label{sec:theory}
In this section we consider  the precession of  the disc induced by the binary companion and how torques between the disc and an inner planet
and subsequently between two planets can enable the system as a whole to precess approximately as a rigid body.

\subsection{Torque acting on the disc due to a binary companion}\label{precto}
 The response of the disc,  with no planets present,  to a binary perturber in an inclined circular  orbit  has been studied by  eg. \cite{Pap1995} and \cite{Lar1996}.
Adopting  their  formalism,  we obtain  the precession frequency  of a disc  with our   surface  density profile.
This is appropriate when  the disc is able to communicate with itself,  either through wave propagation or viscous diffusion,  on a time-scale less than the inverse precession frequency.  
In that case the disc precesses approximately like  a rigid body. This is found to be the case for our simulations (see below).
The precession is retrograde with   period   $2\pi/\omega_0$,  where $\omega_0$ is  given by  \citep[see   Eq. 21 in ][]{Lar1996}
\begin{eqnarray}
 \omega_0= \left( \frac{3 \rmn{G} M_B}{4 D^3}\right) \cos i_B \frac{\int_{R_{in}}^{R_{out}} \Sigma r^3 dr }{\int_{R_{in}}^{R_{out}} \Sigma r^3 \Omega dr}  \label{eq:omega_p}
\end{eqnarray}

We remark that this takes account of only the  quadrupole term in the expansion of its time averaged  perturbing
potential due to the companion.
Neglecting any evolution of the disc surface density profile
 making the approximation of
 extending the disc radial domain to $[0, R_{out}],$ and  adopting Kepler's law such that
 $\Omega^2(R)= {\rmn{G} M_*}/{R^3}$, we obtain 
\begin{eqnarray}
&& \omega_0 = 
 \left( \frac{3 \rmn{G} M_B}{7 D^3 } \right) \cos (i_B) \frac{1}{\Omega(R_{out})}  
\end{eqnarray}
or equivalently
\begin{eqnarray}
&&\frac{\omega_0}{\Omega(R_{out})} =\left( \frac{3 M_B R_{out}^3}{7 M_* D^3 } \right) \cos (i_B) \label{eq:prec_ratio}
\end{eqnarray}
For $M_B/M_*=1$, $D/R_{out}=5$ and $i_B=\pi/4$, Eq. (\ref{eq:prec_ratio}) gives $\omega_0/\Omega(R_{out})=0.0024$. 
Therefore, the condition for sound to propagate throughout the disc during a precession time, namely  $H/R > \omega_0/\Omega(R_{out}),$ is well satisfied for our simulations. 
This is the condition for approximate rigid body precession when the disc is in the bending wave regime.
This occurs for the viscosity parameter $\alpha_{SS} <  H/R,$
which should be applicable here  \citep[see  ][ and Paper 1]{Lar1996}.
Numerically, we find  that  $\Omega(R_{out})=2.22 \times 10^{-9} \rmn{s^{-1}}$. and $\omega_0=0.0024 \times 2.22 \times 10^{-9} \rmn{s^{-1}} =5.3 \times 10^{-12} \rmn{s^{-1}}.$
The precession period is accordingly,  $T_p=2\pi/\omega_0 \sim 1.19  \times 10^{12}\ \rmn{s} \sim 37709 \rmn{yr}$ or
in 3373 time units.  We remark that  as $\omega_0\propto D^{-3},$
this implies that binaries with separations up to $400$~AU could have significant effects over typical protoplanetary disc lifetimes.

\subsection{Torques  between  the different components of the system and the maintenance of approximate coplanarity}

When  planets are present  together with  the disc, 
 torques  act  between them, as well as between the  planets and  the disc. 
 The torque  due to  the  binary companion acts mainly on  the outer parts of the disc.  Its effect may be communicated to planets in  the inner disc regions
 through the action of these  additional torques. 
In appendix \ref{ap:precession}, we derive expressions for 
 them  together with  the  precession frequencies they induce,
assuming that the angular momentum vectors of the different components are approximately aligned.
 Here, we  use these expressions  to estimate  whether 
the precessional torques are large enough in magnitude to allow approximate coplanarity
  to be maintained. As indicated in appendix  \ref{ap:precession},  this requires that the magnitudes  of the  characteristic 
   precession frequencies that could be induced by the additional torques
  be significantly larger than the magnitude of the precession frequency induced by the binary companion.

The retrograde precession rate induced in  the orbit $M_{p,2},$ with radius $r_{p,2},$  by the disc 
can be estimated from (\ref{w2pd}) of appendix  \ref{ap:precession}  to be 
\begin{eqnarray}
\frac{\omega_{p2}}{\Omega_0}&\hspace{-3mm} \sim&\hspace{-3mm} \frac{3 M_D}{4M_{\odot}}\frac{r_{p,2}^{3/2}R_0^{3/2}}{ (R_{out}R_{in,d})^{3/2}} ,
\label{w2pd0}
\end{eqnarray}
where we have expressed the precession frequency in units of the orbital frequency, $\Omega_0,$
at the reference radius $R_0$ used to define our dimensionless units. 

For our simulations we adopt $R_0= R_{in,d}= 5\ \rmn{AU},$ $R_{out}=20\ \rmn{AU}$, $ M_D= 10^{-2}\ \rmn{M}_{\odot}$
and  a simple model
with the presence of a gap and the planet residing in an inner cavity with
$r_{p,2}= 0.8R_{in,d}.$ We then find $\omega_{p2}/\Omega_0\sim 6.7\times 10^{-4}.$
 In addition, direct numerical evaluation using equations (\ref{Torque4}) and (\ref{w2p}) 
indicates this  estimate should be increased by a factor of $2.3.$
 For our simulations we have $\omega_{0}/\Omega_0 < \sim 4\times 10^{-4},$
thus the  above discussion  indicates  that $\omega_{p2}$ may significantly exceed $\omega_0,$  
and so  torques acting between the planet and disc are potentially able to maintain uniform precession.

In the case of a planetary system, the innermost planets, being more distant from most of its  mass distribution,
 are less affected by disc. However, uniform precession may be maintained through torques due to other planets.
We now suppose $M_{p,2}$ is acted on by an exterior planet $M_{p,1}.$ 
Using equation (\ref{w2pp}) we estimate the precession frequency induced by $M_{p,1}$ on $M_{p,2}$ to be
\begin{eqnarray}
\frac{\omega_{p2}}{\Omega_0}&=& \frac{M_{p,1}r_{p,2}^{1/2} R_0^{3/2}}{4M_{\odot}r_{p,1}^2}  b^{1}_{3/2}(r_{p,2}/r_{p,1}).
\label{w2pp0}
\end{eqnarray}
Taking  $ M_{p,1}= 2\times 10^{-3}M_{\odot},$   $r_{p,1}=R_{0}$ and  $r_{p,2}/r_{p,1}=0.6,$ corresponding to proximity to the 2:1 resonance, we obtain
 $\omega_{p2}/\Omega_0 = 1.25\times 10^{-3}.$
Thus we  again conclude that torques induced by $M_{p,1}$ on $M_{p,2}$ are potentially able to maintain uniform precession.

\section{Simulation results} \label{sec:1pl}

\subsection{Single planet in the presence of a gas disc}\label{sec:1p2}

\begin{figure}
\centering
\includegraphics[width=7cm]{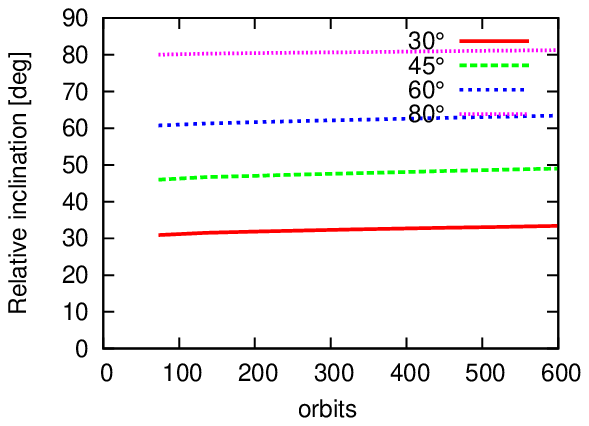}
\includegraphics[width=7cm]{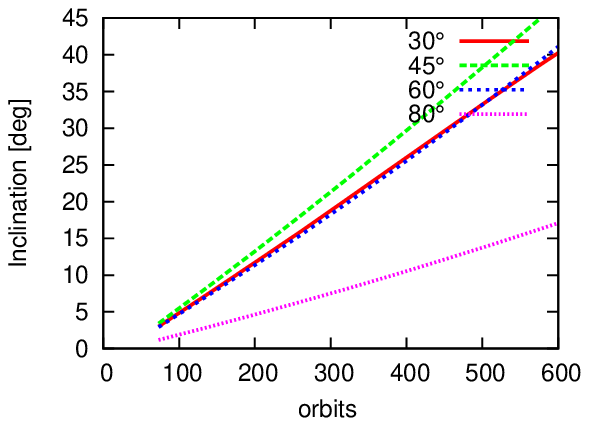}
\includegraphics[width=7cm]{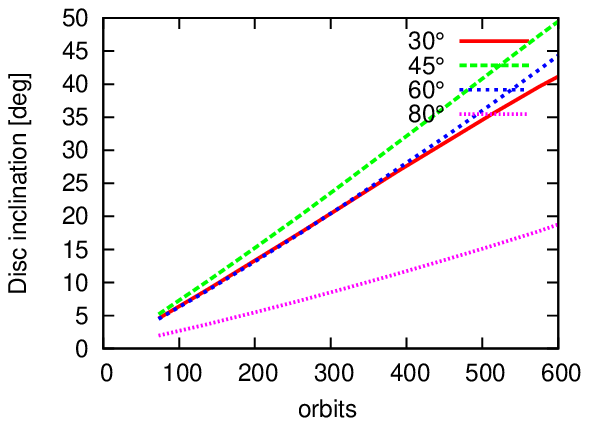}
\caption{Inclination of the angular momentum vector of the planet with respect to that of the binary 
companion (top panel),
 inclination of the planetary orbit and the  disc with respect to the $(x,y)$ plane, the original
disc midplane,  as functions of time  ( middle and bottom panel respectively)
 for different  inclinations $i_B$ of the binary orbit.}
\label{fig:1p_inclination}
\end{figure}

We study the evolution of a single planet of mass $2\ \rmn{M_J}$ in the presence of a gas disc for different initial inclinations, $i_B,$ of the binary companion
to the initial midpllane of the disc.
We show the dynamical evolution of the planet-disc system for  $i_B = 30^{\circ}$, $i_B= 45^{\circ}$, $i_B= 60^{\circ}$ and $i_B = 80^{\circ}.$ 
Simulations were run for  600 orbits because beyond this point the inwardly migrating planet started to become affected by particles  that accumulated in  the innermost 
 boundary region
in the vicinity of  the central star. 
 We plot a variety of   inclination angles as  defined   below.

The orbital inclination of the planet is measured with respect to the $(x,y)$ plane. It is thus given by
\begin{eqnarray}
&i_{p}&= \arccos\left( \frac{{\bf L}_{p,z} }{|{\bf L}_{p}|} \right)\ , \label{eq:i_p} 
\end{eqnarray}
where ${\bf L}_{p}$ is the angular momentum vector of the planet and ${\bf L}_{p,z}$ its $z$-component.
The relative inclination, $i_{rel,p}$, of the planetary orbit to the binary orbit  is then obtained from
\begin{eqnarray}
&i_{rel,p}&= \arccos\left( {\frac{{\bf L}_{p} \cdot {\bf L}_{B}}{|{\bf L}_{p}| \cdot |{\bf L}_{B}|}}\right), \hspace{2mm} {\rm  where} \\
&{\bf L}_{B}&=M_B ({\bf r}_B \times {\bf v}_B),\nonumber \label{eq:rel_i_p}
\end{eqnarray}
 is the angular momentum vector of the binary star which does not vary with  time on account of the orbit being fixed.

 The disc inclination measured with respect to the $(x,y)$ plane  is obtained from
\begin{eqnarray}
\hspace{-1mm}&i_{D}&= \arccos\left( \frac{{\bf L}_{D,z} }{|{\bf L}_{D}|} \right)\ ,  \hspace{2mm} {\rm where} \\
\hspace{-1mm} &{\bf L}_{D}&=\sum_i m_i ({\bf r}_i \times {\bf v}_i)\ ,\, {\rm is \, \,  the\, \, total\, \, angular\,\, momentum} \nonumber \label{eq:i_D} 
\end{eqnarray}
 of the disc obtained
by summing the contributions from  each active  particle and ${\bf L}_{D,z}$ is its $z$ component.

\begin{figure}
\centering
\includegraphics[width=6cm]{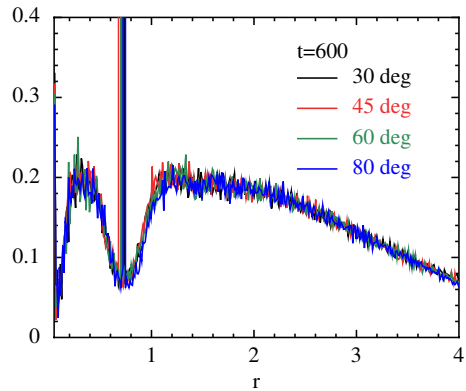}
\caption{Surface density   for  a gaseous disc together with  a $2\ \rmn{M_J}$ planet in the presence of an inclined binary star of mass $1\ \rmn{M_\odot}$
for  different values of $i_B.$  The different curves almost coincide. The spike at the location of the planet is due to gas bound to it.
The indicated  coordinate,  $r, $ is the
 distance to the central star.}
\label{fig:1p_sigma}
\end{figure}

Figure \ref{fig:1p_inclination} shows the  inclination of the 
angular momentum vector of the planet with repect to that of the binary , and  the inclination of the planetary orbit  and  the disc 
with respect to the $(x,y)$ plane as a function of time for $i_B = 30^{\circ}$, $i_B= 45^{\circ}$, $i_B= 60^{\circ}$ and $i_B = 80^{\circ}.$ 
It will be seen that the relative inclination of the planet and binary companion orbits   is almost constant for the entire simulation period. 
Furthermore
the evolution of the  inclination  of the planetary orbit with  respect to the $(x,y)$ plane is very similar
to that of  the disc inclination with  respect to the same  plane for all values of $i_B.$
 This indicates  that the planetary orbit  and disc remain  coplanar during the entire simulation  for all  inclinations of the binary orbit studied. 
In each case the inclinations with respect to the $(x,y)$ plane  are seen to increase approximately linearly with time.
This will be seen to be a consequence of the approximately uniform precession of the disc-planet system.
It means that the system becomes increasingly misaligned with respect to its original plane. The latter could coincide with the equatorial plane
of the central star.

The amount of  inclination increase after a given time is a function of $i_B.$
If this angle is $\beta,$ we have $\sin\beta/2= \sin i_B\sin\omega_0 t/2$
assuming  rigid body precession.   Then from equation (\ref{eq:omega_p}) we see that
 the amount of evolution after a given time
 is initially expected to be proportional to $\omega_0\sin i_B,$ or $\sin{2i_B}$. 
This assumes a linear increase with time, which is expected to be an accurate approximation (to within at most
a few $\%$  as long as the inclination change
  is less than $\sim 50^{\circ},$ as is the case for the results in  Fig. \ref{fig:1p_inclination} at the latest times.
It  is  apparent  that  
the largest inclination change occurs for  $i_B = 45^{\circ}, $ attaining
  $\sim 50^{\circ}$ by the end of the simulation.
We remark that this implies a precession period that is   reasonably consistent with equation
 (\ref{eq:prec_ratio}) (see also the discussion below and in Section \ref{precto})
confirming our simple modelling approach.

 In confirmation of the above discussion,
 the inclinations for $i_B = 30^{\circ}$ and  $i_B = 60^{\circ}$  are  seen to  increase  in almost the same
way as expected.
Furthermore, the  inclination changes  approximately three times slower for  $i_B = 80^{\circ}$ as compared to  $i_B = 45^{\circ}.$
This  is approximately equal to the expected value of  $(\sin {160^{\circ}})^{-1}.$

\begin{figure}
\centering
\includegraphics[width=7cm]{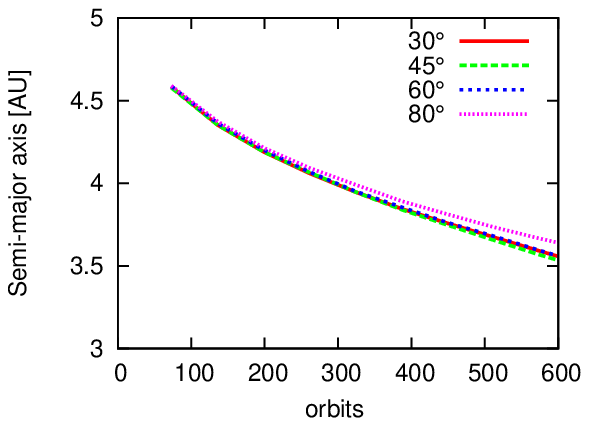}
\includegraphics[width=7cm]{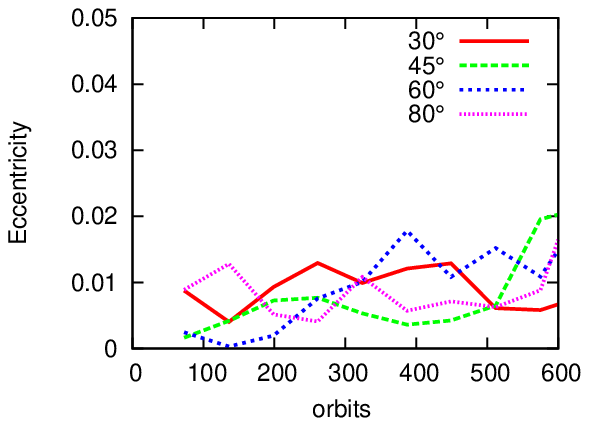}
\caption{
The evolution of the planetary semi-major axis $a_p$ (upper panel) and eccentricity $e_p$ (lower panel)
 for the simulations shown in Fig. \ref{fig:1p_sigma}.
Little variation with $i_B$ is seen.}
\label{fig:1p_figs}
\end{figure}

Since the planet  and the disc remain approximately  coplanar  for all inclinations $i_B$ of the binary orbit,
the mass surface density profile of the disc is also very similar in all inclination cases, as is clearly illustrated  in Figure \ref{fig:1p_sigma}.
Notably,  the structure of the gap induced by the planet hardly varies.

Fig. \ref{fig:1p_figs} shows the evolution of the semi-major axis $a_p$ and eccentricity $e_p$ of the planet. 
Since for all studied $i_B$, the planetary orbit and  disc are approximately coplanar with essentially   the same surface density 
distribution  and  gap profile, the evolution of the semi-major axis is  expected to be the same for all $i_B.$
This is confirmed in Figure \ref{fig:1p_figs}. The planet is seen to migrate inwards from $5$~AU to $3.5$~AU during the course of the simulation.
In addition no significant eccentricity is generated in  the planetary orbit
which is a common feature of the simulations presented here.

\begin{figure}
\centering
\includegraphics[width=7cm]{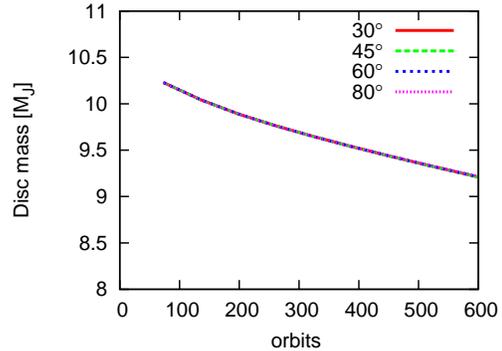}
\caption{The evolution of the disc mass over 600 orbits. The  behaviour for  different inclinations is  almost identical.}
\label{fig:1p_M_d}
\end{figure}

\begin{figure}
\centering
\includegraphics[width=7cm]{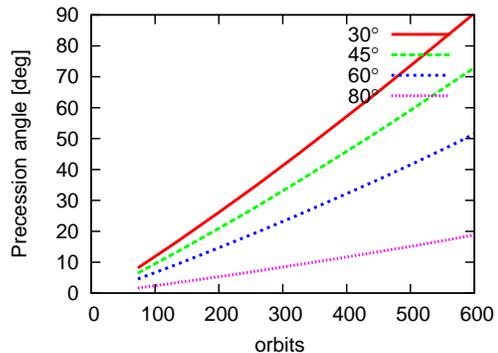}
\caption{The evolution of the precession angle $\beta_p$  for the simulations shown in Fig. \ref{fig:1p_sigma}.
These are approximately consistent with a uniform precession rate $\propto \cos i_B.$ }
\label{fig:theta}
\end{figure}

The disc mass shown in Figure \ref{fig:1p_M_d} decreases from 10 $\rmn{M_J}$ at the simulation start to roughly 9 $\rmn{M_J}$ after 600 orbits. This  evolution is typical  for all simulations shown in this paper.

Fig. \ref{fig:theta} shows the evolution of the precession angle of the disc angular momentum vector, ${\bf L}_D,$ 
around the angular momentum vector of the binary star, ${\bf L}_{B},$ for different  $i_B$.
The precession angle of the disc,  $\beta_p(i_B),$  regarded as a function of $i_B,$ is defined as
\begin{eqnarray}
\cos \beta_p &= &\frac{{\bf L}_D \times {\bf L}_B}{|{\bf L}_D \times {\bf L}_B|} \cdot {\bf u}\, 
\end{eqnarray}
 \citep[See also the definition of $\beta_p$ in][]{Lar1996}. For ${\bf u}$, any fixed unit reference vector in the $(x,y)$  plane could be used.
 In our case, we made  the following choice:
\begin{eqnarray}
 {\bf u}=\frac{{\bf L}_{D,0} \times {\bf L}_B}{|{\bf L}_{D,0} \times {\bf L}_B|}\ ,
\end{eqnarray}
where  ${\bf L}_{D,0}$ is  the total disc angular momentum at time $t=0.$ 
Assuming the presence of the planet can be neglected
and that we have  rigid body precession induced by the time averaged gravitational potential of the  binary companion, the
  precession rate  is expected to be 
 proportional to $\left(\cos{i_B}\right)^{-1}.$  
After 600 orbits, the precession angles are found to be 
 $\beta_p(30^{\circ})=90^{\circ}$,  $\beta_p(45^{\circ})=72^{\circ}$, $\beta_p(60^{\circ})=50^{\circ}$, $\beta_p(80^{\circ})=20^{\circ}$.
Thus $\beta_p(30^{\circ})/\beta_p(45^{\circ})=1.25$ which is in good agreement with the theoretical expectation of $\cos {30^{\circ}}/\cos{45^{\circ}}\cong 1.2$. 
Similarly,  $\beta_p(60^{\circ})/\beta_p(80^{\circ})=2.5$ as against the  expected value $\cos {60^{\circ}}/\cos{80^{\circ}}\cong 2.9$.
Also the comparison between the two extreme cases $\beta_p(30^{\circ})/\beta_p(80^{\circ})=4.5$ shows a satisfactory agreement with the theoretical expectation of $\cos {30^{\circ}}/\cos{80^{\circ}}\cong 5$.

 But note that the results illustrated in Fig. \ref{fig:theta} indicate that the precession frequency increases very  slowly
with time. This evolution is  an expected  consequence of  the accretion,  together with the outward expansion,  of disc material.  Even though
the disc mass distribution changes with time, the system continues to undergo approximate rigid body precession.

\subsection{Influence of the gas disc and a single planet upon each other}

\begin{figure}
\centering

\includegraphics[width=7cm]{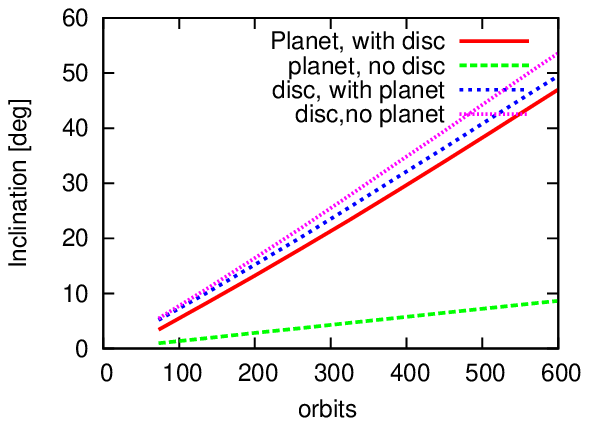} 
\includegraphics[width=7cm]{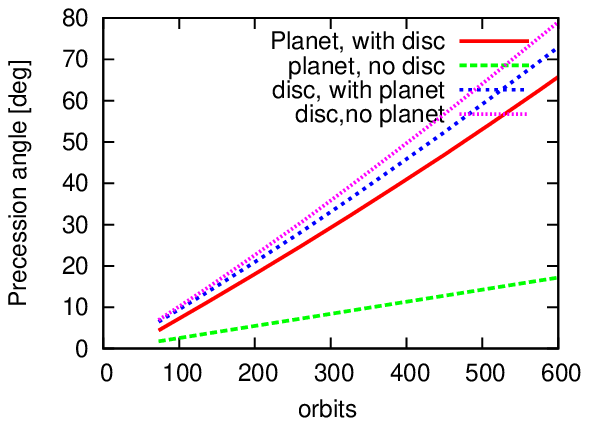}

\caption{Study of the coupling  between disc and planet for $i_B=45^{\circ}$.
The evolution of  the inclinations of  the disc and planetary orbit  with respect to the $(x,y)$  plane
for  three simulations are shown in the upper panel. See the text for the allocation of the different  curves 
The evolution of the corresponding precession angles is indicated in the lower panel.}
\label{fig:effects}
\end{figure}

In the simulations described above,  the planet and  disc are coupled to each other
 such that the planetary orbit and disc midplane  remain approximately
  coplanar independently of  which $i_B$ we chose.
We  now study the influence of the gas disc and the  planet on each other in more detail for $i_B=45^{\circ}$.

In the upper panel  of Fig. \ref{fig:effects}, we show the evolution  of  various inclinations for three different simulations.
The first of these was  the  simulation  of the planet,  disc  and binary companion  described above.
The second was a simulation with  only the disc and binary companion,  and the third was a simulation with only the planet and binary companion.
Thus the planet  was  omitted in the second simulation and the disc was omitted in the third simulation.
 The red line  shows  the  evolution of the planetary  orbit inclination  with respect to the $(x,y)$ plane
  in the first simulation,   while the blue dotted line shows  the  corresponding inclination of the disc in the same simulation.
   Both show a very  similar evolution on account of the strong coupling between the planet and disc..
The purple dotted line shows the evolution of the disc inclination for the second simulation without the planet and the
 green line shows the evolution of the planetary orbit inclination for the third simulation for which the disc was omitted.
It is  seen clearly  that the disc inclination evolution without a planet is similar  to the disc
inclination  evolution with a planet.
 In contrast,  the evolution of the planetary orbit  inclination in the absence of the  gas disc is significantly different to  
 the  evolution that occurs when the gas disc is present.
 In fact the evolution in the latter case is very much slower because the
  interaction of the planet  with the binary companion when the disc is absent 
    is very much weaker than the interaction 
  of the disc with the binary companion independently of whether the planet is present.
 These results clearly indicate the strong coupling between planet and  disc when both are present.

This can be confirmed  by considering  the evolution of the  precession angles   shown in the lower panel  of Fig. \ref{fig:effects}.
While the evolution for the disc is almost independent of the presence of the planet,
 the evolution for the planet  in the  combined system  differs  significantly from the evolution  
in the absence of the  gas disc.
In the presence of the  disc,  the planet's precession angle after 600 orbits is $\beta_p=65^{\circ}$ while in the absence of the disc, it  has  only  reached $\beta_p=18^{\circ}$
 being  almost four times smaller.

\begin{figure}
\centering
 {{\includegraphics[width=10cm]{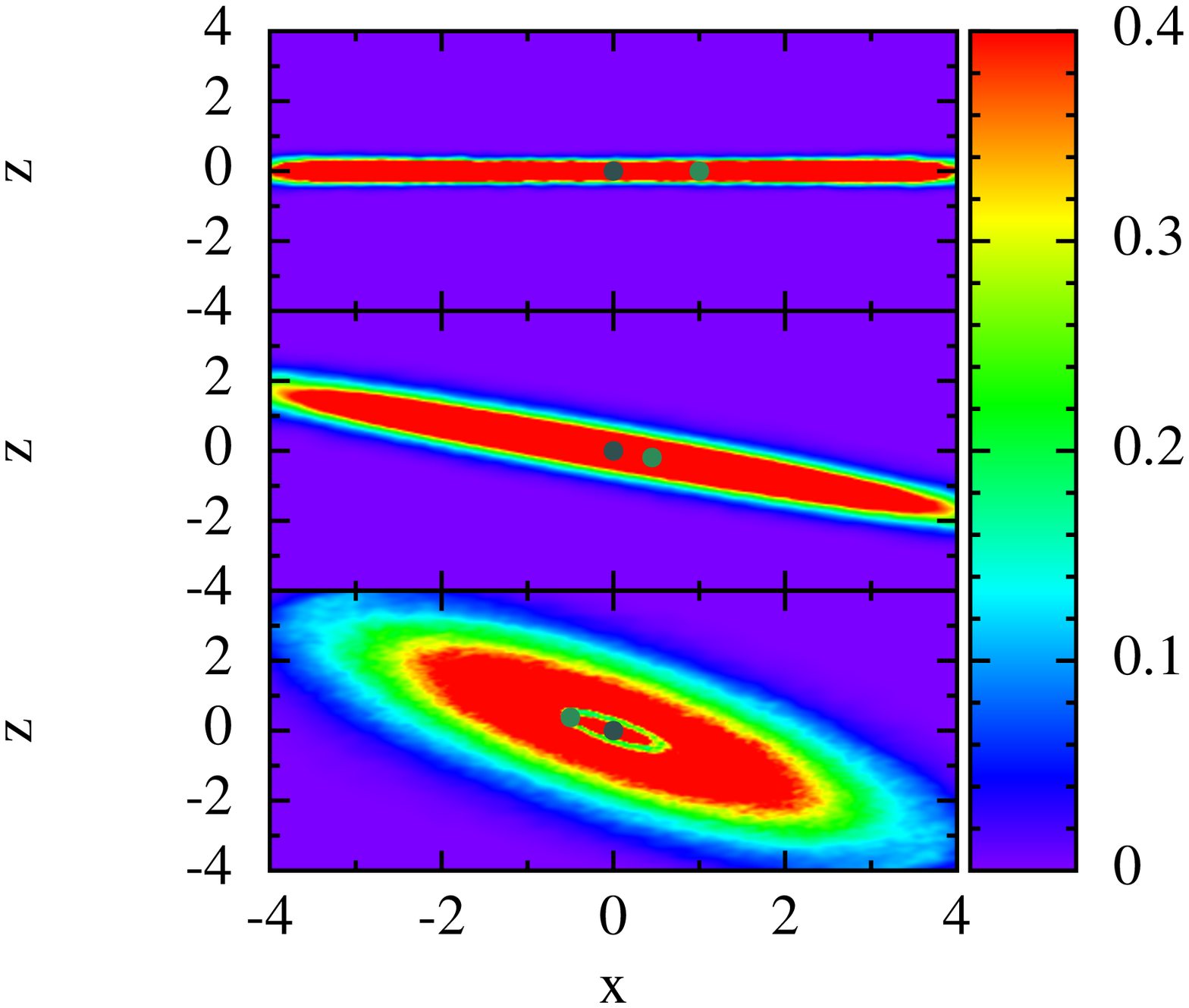} }}%
\qquad
 {{\includegraphics[width=10cm]{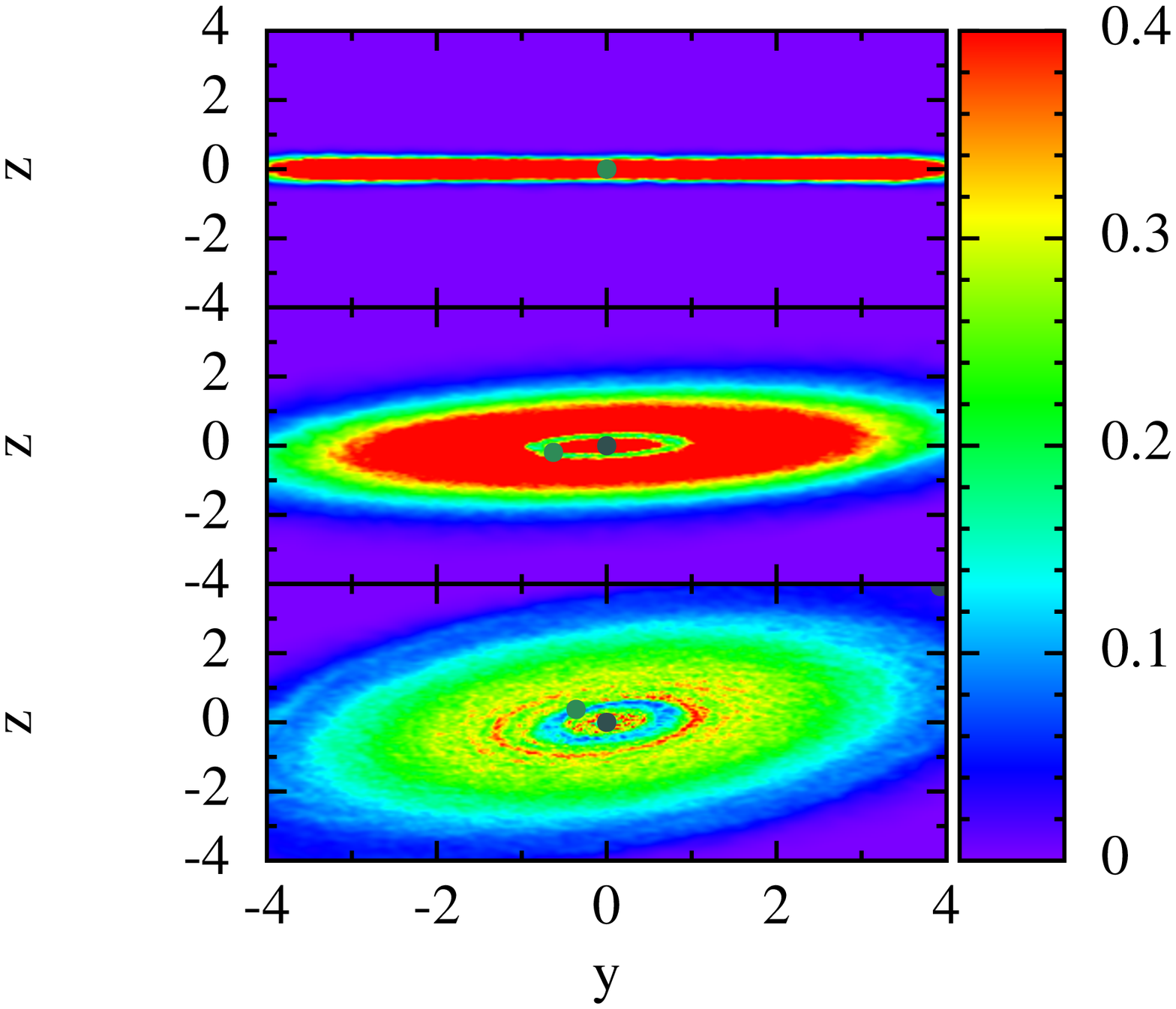}}}%
\caption{  The surface density of the disc on the $(x,z)$ plane  (upper set of three plots) and the $(y,z)$ plane (lower set of three  plots)
for the simulation with disc and planet  illustrated in Fig. \ref{fig:effects}.
In each case the three plots  moving from top to bottom  are respectively at  $t =0,$ $t = 300,$ and $t = 600$ orbits.
The changing orientation of the disc, viewed  from the original fixed  coordinate system, as a function of time, as well as the location of the planet and gap
is displayed. Note that the planet is not seen in the uppermost of the lower set of plots because it is  located on the $y$ axis.
The  unit of length is  $5\ \rmn{AU}$ and the surface density indicated in the colour bar  is given in $\rmn{M_J/(5 \rmn{AU})^2}$.}
\label{fig:effects_rho}
\end{figure}

In order to indicate the evolution of the  geometrical  form of the disc in the first simulation,   we show the  surface density of the disc  projected onto  the $(x,z)$ and  $(y,z)$ planes in Fig. \ref{fig:effects_rho}.
after  $t =0,$ $t = 300,$ and $t = 600$ orbits. These plots indicate only weak warping of the disc as its orientation  changes.
This is  a recurrent  feature of the simulations presented here. The near coplanarity of the planetary orbit is indicated by the fact that the planet is always seen to be close to the disc midplane. 

The above  results   lead to the conclusion that it is the disc which dominates the evolution of the whole disc-planet  system. 
The orientation of the planetary orbit  just follows the disc,  maintaining a  near coplanar orientation while  a gap is maintained inside the disc.

\onecolumn
\twocolumn

\subsection{Simulations with a  1 $\rmn{M_J}$ planet}

\begin{figure}
\centering
\includegraphics[width=7cm]{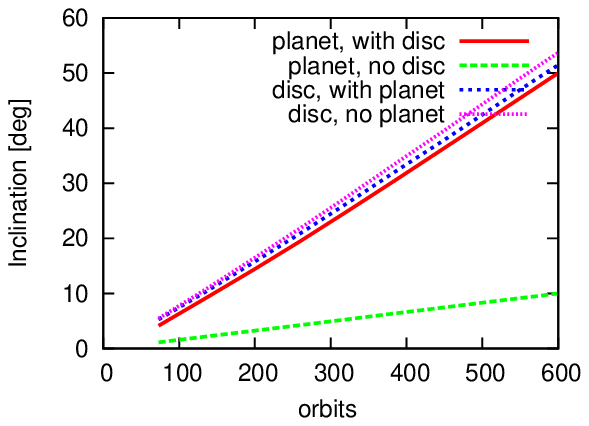}
\includegraphics[width=7cm]{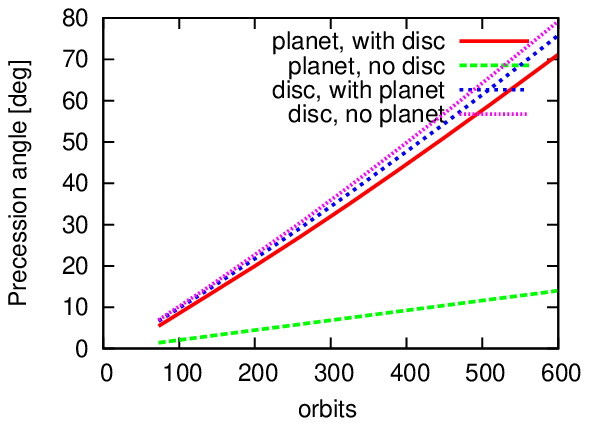}
\caption{As in Fig. \ref{fig:effects} but for a planet of mass  1 $\rmn{M_J}.$}
\label{fig:1M_J}
\end{figure}

The previous simulations incorporated a  planet  of  mass $M_p=2\ \rmn{M_J}.$  In this section,
 we give  results for  a planet of mass 1 $\rmn{M_J},$  showing  that the same conclusions regarding the coupling  of planet and disc  apply.

Fig.  \ref{fig:1M_J} shows the evolution of the inclination  with respect to the $(x,y)$ plane (upper panel) and precession angle (lower panel) of the planetary orbit  and disc 
 for a simulation for which  both components were  included and  two simulations for which  each   component was considered alone.
 Again, the  evolution of the disc orientation   is not significantly affected by the  planet.
  In contrast, the  evolution of the planetary orbit orientation   and its precession is accelerated by the disc.
Compared to the  $2\ \rmn{M_J}$ case,  the lower mass planet appears to be more tightly coupled to the disc
when that is present.
In general the system  is found to  behave  similarly  to the system with a $2\ \rmn{M_J}$ planet.

\subsection{Adiabatic removal of the binary star}\label{Adi}

\begin{figure}
\centering
\includegraphics[width=7cm]{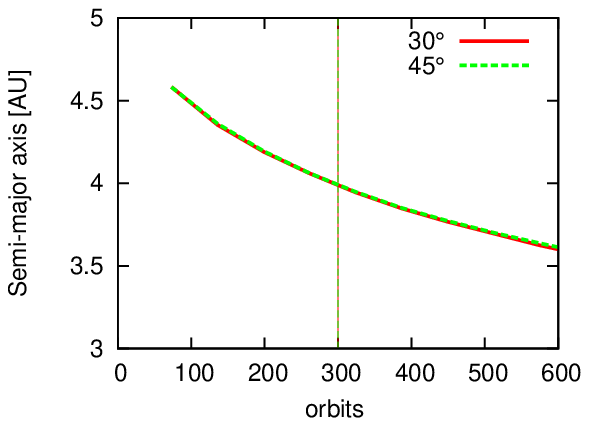}
\includegraphics[width=7cm]{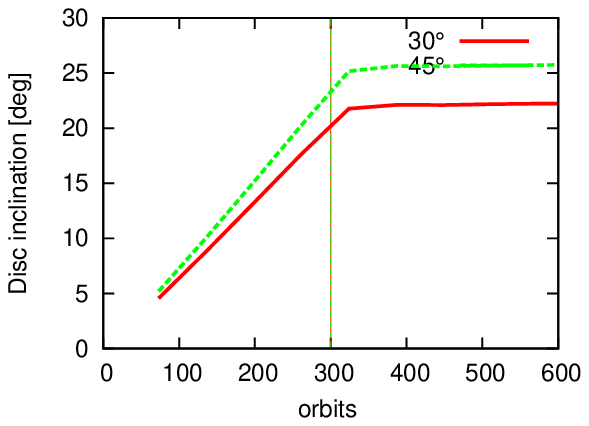}
\includegraphics[width=7cm]{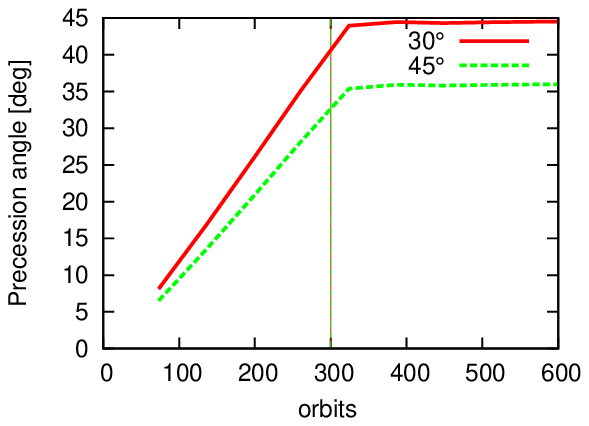}
\caption{  The evolution of the semi-major axis of the planet (top panel), the disc inclination to the $(x,y)$ plane 
(middle panel),  and the disc precession angle   (bottom panel) when  the mass of the binary companion star
is gradually reduced to zero  after 300 orbits. Results for $i_B=30^{\circ}$ and $45^{\circ}$ are shown.}
\label{fig:deactivate}
\end{figure}
When  the binary companion is only present temporarily, as long as the removal process has a long characteristic time scale, as would be expected,  for example,
if it were associated with  an accumulation of stochastic perturbations that might be produced in a star cluster we would expect the disc-planet system to respond adiabatically.
The result could then be found by combining simulations  for fixed binary orbits of the type carried out here.

Although we  consider circular orbits, we remark that our results indicate that the
dominant effect of the companion arises through the  quadrupole term in the expansion of its time averaged  perturbing
potential at large distances. This term takes the same form for an eccentric orbit provided $D$ is replaced 
by   $A_B(1-e_B^2),$ where $A_B$ and $e_B$ are  its semi-major axis and eccentricity  respectively \citep[eg.][]{Yoko}.
Accordingly we expect our results to have a wider application.

In order to study this, we ran  test simulations in which the mass of the binary star was slowly  reduced over a period of $50$ orbits.
for $i_B=30^{\circ}$ and $i_B=45^{\circ}.$  Although this removal time exceeds the characteristic dynamical time in the disc, it is  similar to
  but not long compared to the binary orbital period. Even so the disc-planet system adjusts adiabatically showing no evidence of impulsive kicks.

In Figure \ref{fig:deactivate}, the evolution of the semi-major axis of the planet, the inclination of the disc to the $(x,y)$ plane
and the disc precession angle  is shown.
The vertical lines markxs the starting time of initiation of the  reduction of the binary companion mass.
 For both $i_B=30^{\circ}$(red lines) and $i_B=45^{\circ}$(green lines), this  is after $t=300$ orbits.
The semi-major axis of the planet decreases smoothly  as a result of  inward migration.
 The binary companion mass reduction does  not affect 
 this plot since the relative orientation of the planet and disc stays constant throughout.
In contrast to the semi-major axis, the disc inclination to the $(x,y)$ plane and
 its precession angle are affected by the removal of the binary star.
 When the binary companion has been  removed, it  no longer exerts a  torque  on the disc.
 As a consequence,   both
 the disc inclination to the $(x,y)$ plane and its  precession angle become constant. 
The adjustment occurs smoothly indicating an adiabatic process.

\section{Multiple planetary systems} \label{sec:3pl}

We also study a planetary system composed of three planets, with masses 0.4, 1 and 2 $\rmn{M_J}$. The innermost planet with 0.4 $\rmn{M_J}$ is put at  5 AU. 
The planets are initiated on circular orbits that are coplanar  with the initial disc plane ($z=0$).
The outer two planets are put on circular orbits  resoectively of radii  7.122AU and  12AU.
This  makes  the system  close to  a  2:3:6 resonance, 
being  slightly outside it, so that convergent  migration  would be  expected  to move them towards it.
In practice a 2:1  resonance  
eventually forms between the outermost pair of planets. 
However,  the innermost pair does not attain an orbital  resonance. After migration, the planetary system  attains
a state that is  quite close to instability  that 
can be  induced  when the system is perturbed. This is seen when the  binary companion is introduced (see below).

\subsection{Comparison of results obtained  with  SPH and  a grid based code}

\begin{figure}
\centering
\includegraphics[width=7cm]{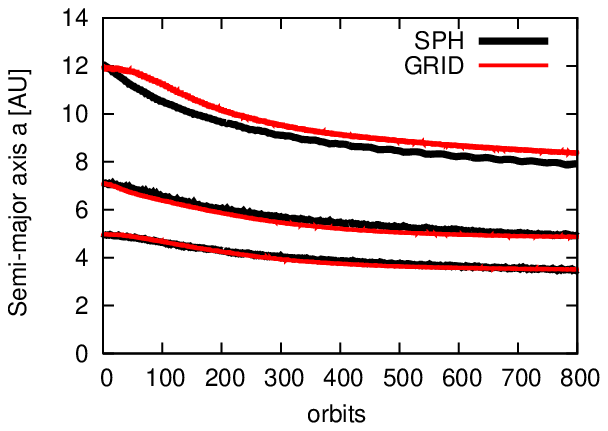}
\includegraphics[width=6cm]{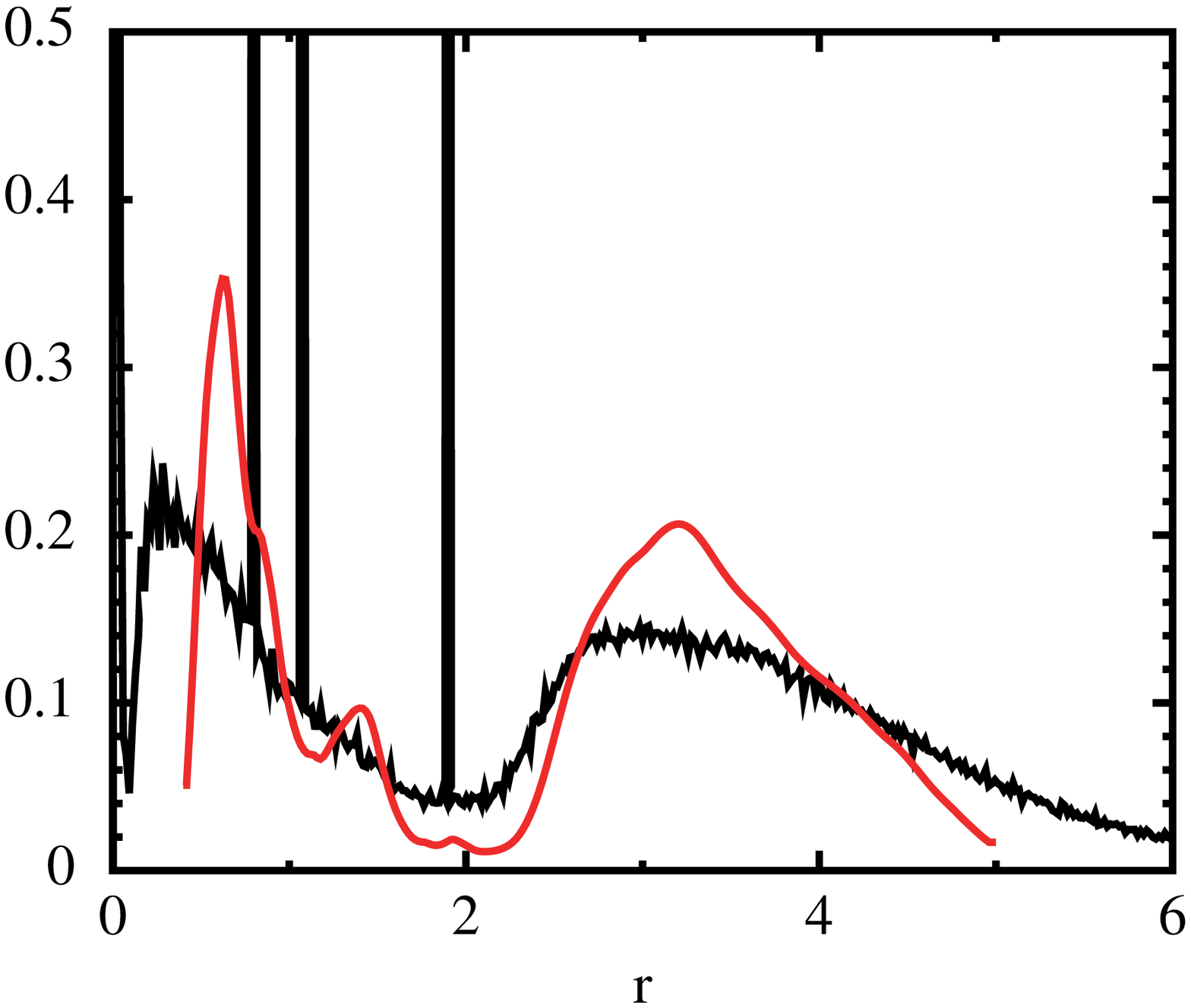}
\includegraphics[width=6cm]{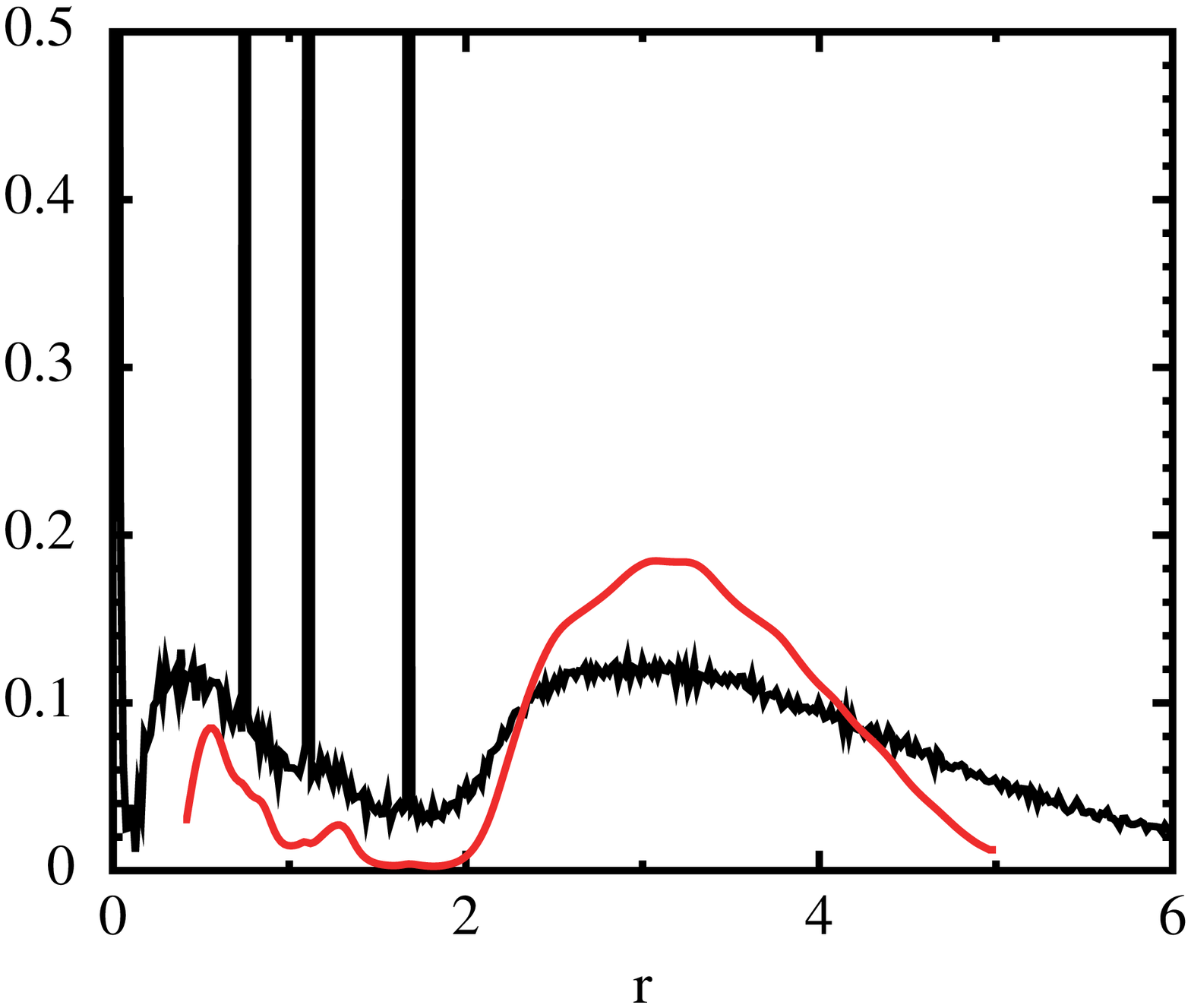}
\caption{Comparison between SPH (black) and grid code (red lines). The upper panel shows the semi-major axes 
of the 3 planets embedded in a gas disc. The lower panels study the surface mass density profiles
 at two different times (see text). The prominent spikes are due to mass accumulation in the vicinity of the planets.}
\label{fig:3pl_SPH}
\end{figure}

As a first step, we compare the evolution of the three planets using  the N-body/SPH code GADGET2 and  a grid based code  without the inclusion of a binary companion.
We performed  two dimensional simulations using  the grid based code NIRVANA 
  \citep[see eg.][ and references therein]{Nel2000}.  Because they are of necessity 2D, these simulations can only be carried out for 
coplanar systems  with no relative orbital  inclinations.
We adopted  a 500$\times$576  equally spaced grid in $(r, \phi),$  finding that reducing
the number of grid points in each direction by a factor of two did not result in significant changes over run times.
An issue is that NIRVANA operates with fixed radial boundaries,  which may be taken to be either rigid or open.
These conditions do not  mimic  the behaviour of the SPH code which allows free outward expansion at large $r,$
and a partial accumulation with some accretion near $r=0.$ We found that  open boundary
conditions  resulted in too much material being lost and instead used partially open boundary conditions
which reduced  outflow speeds by a constant factor of $0.2$ for the inner boundary,  and $0.7$ for the outer boundary,
 as compared to the fully open condition. We remark that the amount of material in the inner regions of the disc is sensitive to these
 conditions and this can affect the inward migration of  giant planets \citep[see eg.][]{Cri2008}.
 Given these complications, nonetheless we  confirm that the very different numerical procedures lead to characteristically similar outcomes.
For the NIRVANA simulations
the inner and outer boundary radii were taken to be $0.4$ and $5$ internal units respectively
The softening parameter for the planet potentials was taken to be $0.6H.$
The  initial surface density profile,  initial planet parameters and circumprimary  disc aspect ratio were 
chosen to be the same as for the SPH code as indicated above.  The kinematic viscosity was taken to be constant with a scaling such that $\alpha_{SS} =0.006$
at a distance of one internal unit from the central star. This value  is somewhat smaller than the value inferred from ring spreading tests.
 However, we found that  a run   for which the viscosity was increased by a factor of three  led to very similar migration  of the planets.
 In these simulations  this depends on the distribution of the embedding material
as well as an  effective viscosity which is operating under different circumstances to those in a ring spreading test.

Tests have shown that the mass of the disc is the main parameter  affecting   the migration rate of the planets.
In this context we note that the masses  of the two components (disc and planets)  contained  within a radial scale characteristic of the outermost planet's  orbital radius are initially comparable.
The local disc mass  subsequently decreases as mass flows towards the star.  
Since (at least) the outer two planets are on resonant orbits they migrate together with  the disc which becomes increasingly unable to drive
 the migration of the planets due to its small mass compared to the total mass of all the planets.
In addition  material  tends to accumulate in an inner disc interior to the planets which  tends to counteract inward migration driven by the
outermost planet.



In the case of the  SPH code, the disc is allowed to evolve in the presence of the three planets for 100 orbits before planetary migration is switched on. By doing this, we allow the disc to smooth out its initial small-scale fluctuations and we allow the planets to open gaps in the disc. In the plot of the semi-major axis evolution
in the uppermost  panel of  Fig. \ref{fig:3pl_SPH}, the initial 100 orbits are not shown for the SPH simulation, i.e. the simulation starting time shown is the time after the initial 100 orbits when the planets start to migrate.
In contrast, in the  case of NIRVANA,  the planetary migration is activated  from the simulation start.
The plot shows that the evolution of the semi-major axes of the planets  is comparable for both codes.
 Only for the outer  most massive planet,  is there a small difference between  the two simulations. 
The reason for this could be the depth of the gap. For a 2 $\rmn{M_J}$ planet,  NIRVANA  shows a deep gap while the SPH code is not able  to produce such  a deep gap.
The excess of material in its neighbourhood  would be expected to lead to a slightly faster inward migration of the massive planet  during the early stage of the simulation. 

The  plots  in the middle and bottom panels of  Fig. \ref{fig:3pl_SPH} compare the  surface densities after 280  and 560 orbits respectively.
 The black lines show  the SPH results while the red lines  shows the  results from NIRVANA.
The surface density profiles  reveal  some differences.
The fixed outer boundary for NIRVANA  is at 5 internal length units (25 AU) beyond which there is no gas. 
 In contrast, the SPH code allows the gas disc to expand  freely with the outer disc edge   becoming  increasingly  smeared out.  In addition
the gaps associated with the two most massive planets,  of 1 and 2 $\rmn{M_J}$ respectively, are deeper for NIRVANA  than for  the SPH case. 
This is a known characteristic difference between  grid based  and SPH   codes  and it  has  been noted  by previous authors \citep[e.g.][]{Val2006}.
The inner disc boundary is treated differently by the two codes. In the SPH case,  particles enter the accretion region around the star, some of which   are accreted and then removed from the simulation.
The inner boundary of the grid based code is partially open with material passing through it being lost. This was taken to be at  $0.4$ internal units (2AU) for numerical convenience.
The consequence is that there is more accumulation of material in the central regions in the  SPH simulation which can eventually affect the inwardly migrating planets.

\subsection{  The effect of an inclined binary orbit}

\begin{figure}
\centering
\includegraphics[width=7cm]{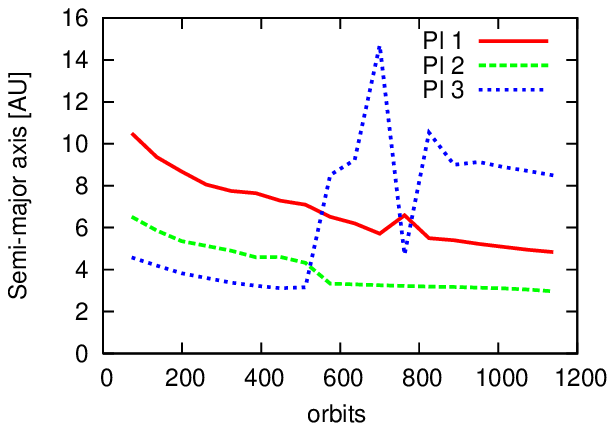}
\includegraphics[width=7cm]{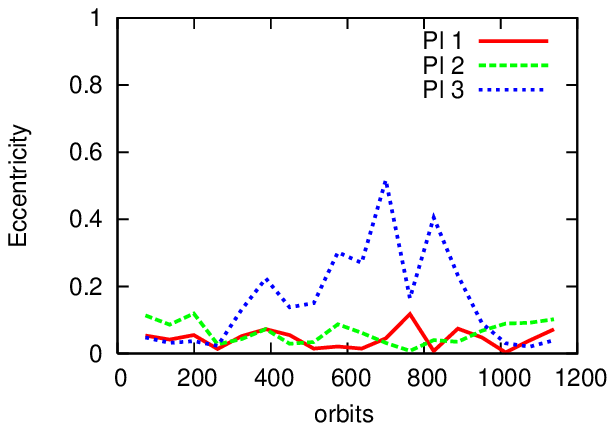}
\includegraphics[width=7cm]{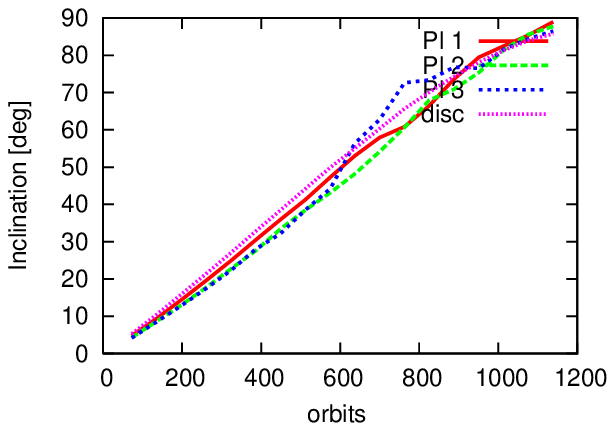}
\caption{Evolution  of a 3 planet system embedded in a gas disc and perturbed by a binary companion with $i_B=45^{\circ}$ and $D=100\ \rmn{AU}$.
The semi-major axes, eccentricities and inclinations with respect to the $(x,y)$ plane of the three planets are shown in the top, middle and bottom panels 
respectively.  The corresponding inclination of the disc is also shown in the bottom panel. 
The initially outermost, central and innermost planets are labelled by Pl 1, Pl 2 and Pl 3 respectively.}
\label{fig:3pl_f}
\end{figure}
We now describe a SPH  simulation starting with the same parameters as that described in the previous section
but now including a $1M_{\odot}$ binary companion in circular orbit with $i_B = 45^{\circ}.$
Results are presented in Fig. \ref{fig:3pl_f}. 
In this simulation the planetary system is  ultimately unstable. Instability starts after $500$ orbits and
leads to close encounters between the planets.
The interactions  produce a new radial ordering of the planets. 
The innermost planet with $M_p=0.4\ \rmn{M_J}$ is scatterred outwards to become the outermost planet.
 After an  additional scatterring, the three planets  stabilize in near circular orbits  after $\sim 900$ orbits.
Although the planetary system is unstable, its  inclination with  respect to the  $z=0$ plane 
 and its  relative inclination with   respect to the binary companion,  remain  coupled to  the disc as for the single planet cases
throughout the evolution. 
\begin{figure}
\includegraphics[width=7cm]{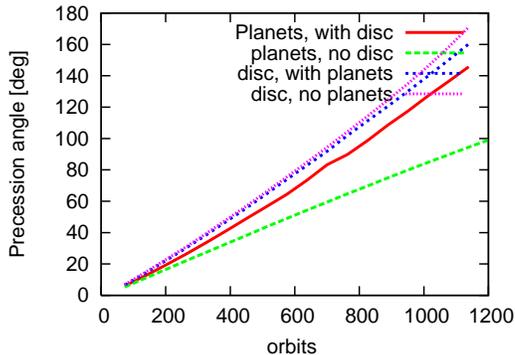}
\caption{Precession angles  for the components  of 
 the simulation shown in Fig. \ref{fig:3pl_f} together with precession angles obtained from simulations where the disc and 
planetary system were included individually without the other component.
As in the latter case migration did not occur the planetary system did not undergo an instability.}
\label{fig:3pl_precession}
\end{figure}
The evolution of precession angles  of the components of the 
 planetary system  are shown in Figure \ref{fig:3pl_precession}.
This reveals the same characteristic evolution as previous simulations with just one planet.
 In the presence of the gas disc, the precession of the planetary angular momentum vector occurs  significantly faster than it does in 
 the absence of the  gas disc.  The observed  precession of the disc with the planets is almost identical to that
 observed when the planets are absent.

\section{Summary and Discussion} \label{sec:concl}

We have performed SPH-simulations in order to study the influence of a  stellar binary companion on a circumstellar disc and massive planets that may orbit  therein.
The gas disc and the planets were initiated on coplanar  circular orbits around the central star.  We studied  four inclinations  of the plane of the binary
orbit  to the initial disc midplane.  These were    $i_B=[30^{\circ}, 45^{\circ},60^{\circ} {\rm and}\,  80^{\circ}]$.
 The masses of the central star and binary companion were taken to be $1\ \rmn{M}_{\odot}.$
 We considered planets of mass $2\ \rmn{M_J}$ and $1\ \rmn{M_J}$  in a disc of mass $0.01\ \rmn{M_\odot}$ with outer radial boundary at $20$~AU. 
 The binary orbit was taken to be circular with orbital radius $100$~AU.
 
The central finding was that the planet and the gas disc evolved  maintaining  approximate coplanarity for the full range of $i_B$.
 Apart from the geometric configuration of the orbits, this coplanarity was manifested in diverse ways,  such as the invariance of the inward planet  migration or the gap profile.
 Importantly and quite generally, the relative inclination of the disc and planet in respect to the binary star was constant through the simulations.  
 The enforcement of coplanarity means that large inclination changes between the plane of the disc-planet system and the original disc plane, which might 
 coincide with  the stellar equatorial  plane, can attain large values, especially for large $i_B.$  Such changes could be made permanent if the binary companion is removed. 

  We remark that we have neglected any  precessional or accretion torques acting on the central star which might
 change the orientation of  its equatorial plane.  However, \citet{Bat2012}  showed that precessional torques could be neglected
 for a slowly rotating star when the disc extended down to  the corotation radius.  Furthermore, they are expected to be significantly weaker for a disc with an inner cavity as envisaged here. 
 Accretion torques are also likely to be weak in these circumstances and notably \citet{Bat2013} indicate that these depend on the form of the stellar magnetic field and do not need to act in a way that would tend to make the system align. 
  
  Although the specific choice of simulation parameters relating to the initial disc size and binary orbit were chosen largely for numerical convenience,
  we point out that the results can be extended to apply to other configurations in a number of ways.  Apart from  being able to apply the usual scaling of lengths and times that leave
  gravitationally interacting systems invariant, we remark that the  important  perturbation component  from the binary companion that is responsible for  precession and orbital orientation changes 
  is the time averaged quadrupole term that is dominant  for large orbital separation. This term takes the same form regardless of the eccentricity of the orbit,
  provided $D$ is replaced by $A_B(1-e_B^2),$ where $A_B$ and $e_B$ are the semi-major axis and eccentricity of the binary orbit respectively.
  Thus, as long as the quadrupole term is dominant we expect similar results. In addition, we estimated the precession period
  from our  simulations  to be  $3.8\times 10^4(20{\rm AU}/R_{out})^{3/2}(D/100{\rm AU})^{3/2}\ \rmn{yr}$ (see Section \ref{precto}). 
  Thus the same precession period is obtained if $R_{out}$ and $D$ are increased by the same factor.  Although the disc is larger, the dynamics  
  in the inner region should not be affected as long as internal communication in the disc allows quasi-rigid body precession.
  Note too that if we increase $D$ without increasing $R_{out},$ the precession period increases making the maintenance of coplanarity even easier.

  In summary, the main factors affecting the maintenance of coplanarity are the ability of the disc to communicate within a precession period 
  and the ability of the components to exert strong enough precessional torques. These can be estimated from equations (\ref{Torque4})-(\ref{w2pp})
  and will be seen to be determined by the disc and planet parameters within a radial scale defined by the planetary system rather than the outer boundary radius.
  This generates many possibilities for producing significant orbital reorientation through interaction with the binary.
  

Furthermore  we also ran test simulations in order to model the slow removal of the binary companion. By doing this, we noted the adiabatic behaviour of the system.
We found a smooth transition of the system to  one with only a central star.
 Due to the lack of the external torque, the system stopped its precession,  attaining a  constant inclination between the midplane of the disc-planet  system  and the original disc midplane.
This indicates that large orbital orientation changes could be produced through a sequence of interactions with different binary companions as long as the dynamical processes associated with evolution
of the binary orbits is long compared to the characteristic dynamical time.
 


We also studied a multi-planet system with three planets and disc under the influence of a binary companion. 
We performed a comparison between the N-body/SPH code GADGET-2 and  a 2D grid based code  NIRVANA  for the system with no binary companion.
The planets with masses 0.4, 1 and 2 $\rmn{M_J}$ were put on coplanar circular orbits near to a 2:3:6 resonance.
Even though there are significant  differences in implementation  such as the treatment of the boundary conditions  and  the application and behaviour  of 
viscosity,  the form of the migration  of the planets is in  satisfactory agreement.

When   the  planetary system  was embedded in a gas disc in the presence of a binary companion, it was found that inward migration
resulted in the appearance of instability.
 After 500 orbits, the innermost planet   of  0.4 $\rmn{M_J}$  was scattered to  attain the outermost orbit
after which 
the evolution stabilized  with the planets again attaining  near circular orbits.
Throughout this evolution,  the  orbital  inclinations  of all three planets remained strongly coupled to the  disc inclination. 
In this respect the multi-planet system behaved exactly as the single planet system. 

Thus the simulations that we have performed have shown that when the disc is induced to undergo quasi-rigid body precession, as long as this is on a long enough timescale
that internal precessional torques can retain coplanarity. This is maintained when there is a migrating planetary system with gaps in the interior part of the disc.

In this regard,  we remark that \cite{Ghe1993} have conducted an observational survey of T Tauri stars in two neighbouring star forming regions and obtained a binary  frequency  
in the  separation range 16 - 252 AU for T Tauri stars of
$\sim$ 60 \% which is roughly 4 times greater than the binary star frequency for solar-type main-sequence stars. As  this frequency  can be altered  by evolutionary processes within  star
 clusters which can form  and disrupt binary systems,  \cite{Ghe1993} proposed that the high binary star frequency associated with  T Tauri stars is correlated with their young ages  and is  thus an evolutionary effect. However, this  could be crucial  for  a  protoplanetary disc, 
 since as shown in our simulations, an inclined binary companion is able to produce a disc with a high inclination with respect to the stellar equatorial plane 
 in a time  that is short compared to the disc's lifetime.
More recent studies of binaries in young  stellar systems mainly support the results of \cite{Ghe1993}  \citep[e.g.][]{ Koe2001, Kra2007,  Kro2011}.
 Although there is a large uncertainty, observations reveal a frequency of binary systems in stellar clusters of 30 $\sim$ 50 \% \citep{Duc2004, Kra2009, Mar2012}.
Note that this is comparable to the frequency of hot Jupiters with orbits that are highly inclined with respect to the stellar equatorial plane. 
Given the high  abundance of binary stellar systems and the short timescale required for inclination generation, a scenario involving  a binary companion 
with separation  of the order of 100 AU   could
 accordingly  give  a possible explanation for the observed frequency of planets on highly  inclined orbits.

\section*{Acknowledgments}

Xiang-Gruess acknowledges support through Leopoldina fellowship programme  (fellowship number LPDS 2009-50).
Simulations were performed using the Darwin Supercomputer of the University of Cambridge High Performance Computing Service, provided by Dell Inc. using Strategic Research Infrastructure Funding from the Higher Education Funding Council for England and funding from the Science and Technology Facilities Council.

\begin{appendix}
  
\section{Conditions for uniform precession} \label{ap:precession}
We consider the conditions for  a system consisting of a  disc and an interior planetary system to
precess uniformly together while maintaining only a small relative inclination under  perturbation
from an object such as an external binary companion. A complete treatment requires an analysis of the internal dynamics
of disk warping together with  the response of the planetary orbits (eg.  Larwood et al. 1996, Terquem et al., 1998).
However, such a discussion is  complex and as we require only approximate estimates, we shall consider a simplified 
approach
based on estimating whether the magnitudes  of  gravitational torques  between different  components of the system  
can be sufficient to enable it
to precess as a whole while maintaining approximate coplanarity. We  proceed by assuming  the latter situation  occurs
and then consider whether it can be maintained by internal torques.

We adopt a simple model for which only
secular interactions are considered.  Thus  possible effects arising from orbital  resonances are neglected.
We also ignore the effects of density  waves excited through disc-planet interactions, making the assumption
that the disc mass distribution is unperturbed. Nonetheless the torque balance we consider should still occur.
The approach can be applied to situations where planets are within a disc gap or inner cavity
or where there is no disc.

\begin{figure}
\includegraphics[width=7cm]{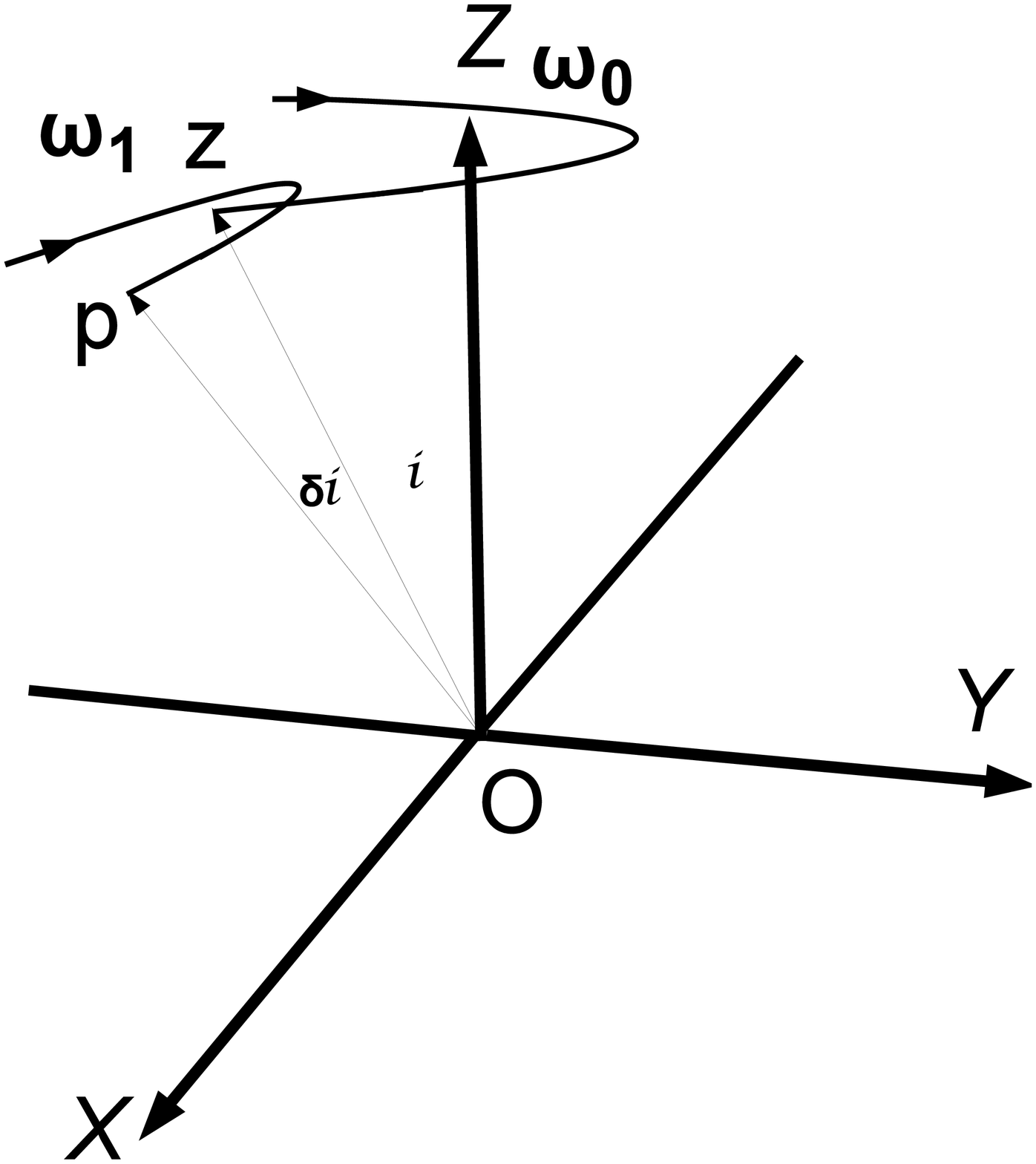}
\caption{Schematic illustration of the precession of the different components of  a  disc-planet system.
A binary companion causes the total angular momentum of the  inclined  system to precess around  the binary orbital angular momentum
vector in the direction $O Z,$ with precession angular velocity ${\mbox{\boldmath$\omega$}_0 }.$
The torque from the binary is too weak to affect the inner planetary system  whose angular momentum vector in the direction $O{\rm p} ,$ is induced  to precess about the total angular momentum vector
of the disc-planet system  in the direction $O{\rm z},$     with precession angular velocity ${\mbox{\boldmath$\omega$}_{1} },$     through  the action of disc-planet,  or planet-planet torques.
The angular momentum vectors  are maintained in the same plane if the components of the precession angular velocity vectors, ${\mbox{\boldmath$\omega$}_0 }$
and  ${\mbox{\boldmath$\omega$}_{1} }$   perpendicular to $O{\rm p} $ are the same.  This requires the inclinations to be adjusted so that 
 $|{\mbox{\boldmath$\omega$}_0 }|\sin(i+\delta i) =  |{\mbox{\boldmath$\omega$}_{1} }|  \sin \delta i.$ This results in a small mutual inclination, $\delta i,$ provided
  $  |{\mbox{\boldmath$\omega$}_{1} }|  \gg     |{\mbox{\boldmath$\omega$}_0 }|.$}
\label{precessioncartoon}
\end{figure}

An  illustration of the type  of  situation we study is given in Fig. \ref{precessioncartoon}.
We consider two   components of the disc-planet system undergoing precession induced by a binary companion.
This is presumed to occur while it maintains an approximate planar structure.
However, only the outer component is directly induced to precess by the binary companion.
The inner component has to adjust a relatively small mutual inclination so that the torque acting on it due to the outer component
causes it to precess at the same rate.  As illustrated in Fig, \ref{precessioncartoon} this requires that the components  of 
the angular velocity of precession  induced by the mutual interaction,
 and the angular  velocity  of precession  of the entire system, that are  perpendicular to the angular momentum  vector of the inner component
be the same.  This is because it is only these components that determine the rotation rate of the latter vector.
 
We remark that many studies (eg.  Larwood et al. 1996)   have shown that a gaseous disc can maintain uniform precession under the action
of torques due to a binary companion,  provided that sonic or diffusive communication  can occur within a precession time scale.
These conditions are satisfied for our simulations  and disc self-gravity    is not required.
Here we focus on the incorporation of an inner planetary system, which on account of depletion of material through gap formation is expected
to require gravitational torques in order to maintain coupling to the outer disc.

As the  precession frequencies of interest  are  much smaller than the characteristic orbital frequency, we average over many orbits
so that the system , incorporating  any planets present, may be considered to consist of a mass distribution.
Assuming orbits to be circular,   both for  planets and in the  disc,
 this may be taken to be a combination of distributions, each of which may be taken to be axisymmetric with appropriate symmetry axes.
 We now discuss the torques between such distributions and the induced precession rates.

\subsection{ Precessional torques  induced by axisymmetric mass distributions}
We consider the precessional torque acting between two mutually inclined  axisymmetric 
mass distributions. The densities associated with these are respectively taken to be $\rho_1({\bf r})$  and  $\rho_2({\bf r}).$ 
We adopt  cylindrical coordinates $(R, \phi, z)$ such that the symmmetry axis asscociated with  $\rho_1({\bf r})$
is the $z$ axis.  The $(x,y)$ plane is assumed to be  a  plane of symmetry such that $\rho_1(R,-z)= \rho_1(R,z).$
The distribution associated with $\rho_2({\bf r})$ has the same symmetry properties but the plane of symmetry
is taken to have an inclination $\delta i$ with respect to the $(x,y)$ plane and accordingly
the axis of symmetry is inclined at an angle $\delta i$ to the $z$ axis. We shall assume the distribution
$\rho_1({\bf r})$ is  located exterior to the  distribution
$\rho_2({\bf r}).$ In addition the $x$ axis is chosen such that the two symmetry planes intersect on it.

There will  be  a  torque of magnitude,  $T_{12x},$  between these distributions acting  in the $x$ direction. This acts to produce precession about their
total angular momentum axis. We shall assume that the outer distribution associated with  $\rho_1({\bf r})$
contains almost all the angular momentum so that the precession will be about the $z$ axis.
The torque magnitude   $T_{12x}$ is readily written down as
\begin{eqnarray}
T_{12x}&=&-\int \rho_2({\bf r})\left( y\frac{\partial \Psi_1}{\partial z} - z\frac{\partial \Psi_1}{\partial y}\right)d^3{\bf r}, 
\label{Torque}
\end{eqnarray}
where $ \Psi_1$ is the gravitational potential arising from  $\rho_1({\bf r})$ and this and similar  integrals below 
are  taken over the  domains occupied  by the relevant  mass distribution.  Substituting for  $ \Psi_1,$  we find
\begin{eqnarray}
T_{12x}&=&-\int\hspace{-2mm}\int G\rho_2({\bf r}) \rho_1({\bf r}')\frac{(zy'-yz') }{|{\bf r} - {\bf r}'|^3} d^3{\bf r}'d^3{\bf r}.
\label{Torque1}
\end{eqnarray}
We assume that $\rho_1({\bf r})$ is confined to the plane $z=0$ forming a distribution appropriate to a flat disc, while $\rho_2({\bf r})$
is confined to a ring corresponding to the time averaged mass distribution appropriate to a planet on a circular orbit.
Then (\ref{Torque1}) reduces to

\begin{eqnarray}
T_{12x}&\hspace{-3mm} =&\hspace{-3mm} -\frac{GM_{p,2}r_{p,2}\sin\delta i}{2\pi}\nonumber\\ 
&& \times \int\hspace{-2mm}\int
 \hspace{-2mm}\int \frac{ R^2 \Sigma(R)\sin\psi \sin\phi d\psi d\phi dR}{( R^2 +r^2_{p,2}  - 2 Rr_{p,2}\cos\alpha ) ^{3/2}},
\label{Torque2}
\end{eqnarray}
where $\cos\alpha = \cos\phi\cos\psi + \sin\phi\sin\psi\cos\delta i,$  $\Sigma$ is the disc surface density,
the mass of the planet is $M_{p,2}$ and its orbital radius is $r_{p,2}.$
When $\Sigma$ is assumed to be localized at radius, $r_{p,1},$ such that the total associated mass is $M_{p,1},$  the torque becomes that acting between  two planets
of masses  $M_{p,1}$ and  $M_{p,2},$ namely
\begin{eqnarray}
T_{12px}&\hspace{-3mm} =&\hspace{-3mm} 
-\frac{GM_{p,1}M_{p,2}r_{p,1}r_{p,2}\sin \delta i}{4\pi^2} 
 \nonumber\\ 
& &\times  \int\hspace{-2mm} \int\hspace{-2mm}\frac{  \sin\psi \sin\phi d\psi d\phi }
{( r_{p,1}^2 +r_{p,2}^2  - 2 r_{p,1}r_{p,2}\cos\alpha ) ^{3/2}}.
\label{Torque3}
\end{eqnarray}
In the limit of small $\delta i, $ which is of interest as it corresponds to close alignment, $\cos \delta i$ may be replaced by unity in the expression for $\cos\alpha$
and the integrals over the angles may be carried out with the result that
\begin{eqnarray}
T_{12x}&\hspace{-3mm} =&\hspace{-3mm} -\frac{1}{2} \pi GM_{p,2}r_{p,2}\sin \delta i  \int  \frac{\Sigma(R) b^{1}_{3/2}(r_{p,2}/R)}{R} dR,
\label{Torque4}
\end{eqnarray}
with  the Laplace coefficient is defined through
\begin{eqnarray}
 b^{1}_{3/2}(x)= &\hspace{-3mm} &\hspace{-3mm} \frac{1}{\pi} \int\frac{\cos\alpha  d\alpha }
{( 1 +x^2  - 2 x\cos\alpha ) ^{3/2}}.
\label{Torque6}
\end{eqnarray}
The corresponding expression for two planets  of masses  $M_{p,1}$ and  $M_{p,2}$ is
\begin{eqnarray}
T_{12px}&\hspace{-3mm} =&\hspace{-3mm} -\frac{GM_{p,1}M_{p,2}r_{p,2}\sin \delta i}{4  r_{p,1}^2  }  b^{1}_{3/2}(r_{p,2}/r_{p,1}).
\label{Torque30}
\end{eqnarray}
Lower bounds for the absolute magnitudes of the  torques can be estimated by replacing $ b^{1}_{3/2}(x)$ by the first term in its series expansion, namely  $3x.$
Thus for a model disc with $\Sigma \propto R^{-1/2}$ truncated at an inner radius, $R_{in,d},$  assumed
to be much less than its  outer truncation  radius,  $R_{out},$ we   integrate (\ref{Torque4})  over $R$ to estimate  the torque exerted by the disc on the planet with
mass $M_{p,2}$ to be
\begin{eqnarray}
T_{12x}&\hspace{-3mm} \sim&\hspace{-3mm} -\frac{3  G M_D M_{p,2}r_{p,2}^2\sin \delta i }{4(R_{out}R_{in,d})^{3/2}} .
\label{Torque50}
\end{eqnarray}
\subsection{Precession frequencies}
The retrograde precession rate induced in  the orbit $M_{p,2}$ by the disc is readily found from
\begin{eqnarray}
\omega_{p2}&=&\frac{ |T_{12x}|}{M_{p,2}r^2_2\Omega_{p,2}\sin \delta i},\label{w2p}
\end{eqnarray} 
where $\Omega_{p,2}$ is the orbital frequency of $M_{p,2}$. 
Using equation (\ref{Torque50}), we obtain the estimate
\begin{eqnarray}
\frac{\omega_{p2}}{\Omega_0}&\hspace{-3mm} \sim&\hspace{-3mm} \frac{3 M_D}{4M_{\odot}}\frac{r_{p,2}^{3/2}R_0^{3/2}}{ (R_{out}R_{in,d})^{3/2}} ,
\label{w2pd}
\end{eqnarray}

In the case of a planetary system,  uniform precession of the innermost planets  may be maintained through torques due to other planets.
We now suppose $M_{p,2}$ is acted on by an exterior planet $M_{p,1}.$ 
Replacing $T_{12x}$ in equation  (\ref{w2p})  by $T_{12px}$  given by  equation   (\ref{Torque30})   we estimate the retrograde  precession frequency induced by
 $M_{p,1}$ on $M_{p,1}$  to be
\begin{eqnarray}
\frac{\omega_{p2}}{\Omega_0}&=& \frac{M_{p,1}r_{p,2}^{1/2} R_0^{3/2}}{4M_{\odot}r_{p,1}^2}  b^{1}_{3/2}(r_{p,2}/r_{p,1}).
\label{w2pp}
\end{eqnarray}

 \end{appendix}

\label{lastpage}

\end{document}